\documentclass{aastex631}

\usepackage{amsmath}   
\usepackage{amsfonts}  
\usepackage{amssymb}   
\usepackage{longtable}  
\usepackage{booktabs}   

\DeclareUnicodeCharacter{02BC}{\textquoteright}

\begin{document}

\title{FALCO: a Foundation model of Astronomical Light Curves for time dOmain astronomy}

\author[0000-0002-0772-6280]{Xiaoxiong Zuo}
\affiliation{National Astronomical Observatories,
  Chinese Academy of Sciences,
  Beijing 100101, People’s Republic of China}
\affiliation{School of Astronomy and Space Science,
  University of Chinese Academy of Sciences,
  Beijing 100049, People’s Republic of China}
\affiliation{National Astronomical Data Center, Beijing 100101, People’s Republic of China
}
  
\author[0000-0002-3143-9337]{Yihan Tao$^{*}$}
\affiliation{National Astronomical Observatories,
  Chinese Academy of Sciences,
  Beijing 100101, People’s Republic of China}
\affiliation{National Astronomical Data Center, Beijing 100101, People’s Republic of China
}
\email{y.tao@nao.cas.cn} 

\author[0000-0003-3250-2876]{Yang Huang$^{*}$}
\affiliation{School of Astronomy and Space Science, University of Chinese Academy of Sciences, Beijing 100049, People’s Republic of China}
\affiliation{Key Laboratory of Optical Astronomy, National Astronomical Observatories, Chinese Academy of Sciences, Beijing 100101, Peopleʼs Republic of China;}

\author{Zhixuan Kang}
\affiliation{Research Center for Astronomical Computing, Zhejiang Lab, Hangzhou 311100, People’s Republic of China}

\author[0009-0000-6108-2730]{Huaxi Chen$^{*}$}
\affiliation{Research Center for Astronomical Computing, Zhejiang Lab, Hangzhou 311100, People’s Republic of China}

\author[0000-0002-7456-1826]{Chenzhou Cui$^{*}$}
\affiliation{National Astronomical Observatories, Chinese Academy of Sciences, Beijing 100101, People’s Republic of China}
\affiliation{National Astronomical Data Center, Beijing 100101, People’s Republic of China
}  

\author[0000-0002-8619-4015]{Jiashu Pan}
\affiliation{School of Engineering, Westlake University,Hangzhou, Zhejiang, 310030, People’s Republic of China}

\author[0000-0001-8011-8401]{Xiao Kong}
\affiliation{National Astronomical Observatories, Chinese Academy of Sciences, Beijing 100101, People’s Republic of China}

\author{Xiaoyu Tang}
\affiliation{Research Center for Astronomical Computing, Zhejiang Lab, Hangzhou 311100, People’s Republic of China}


\author[0000-0003-3474-5118]{Henggeng Han}
\affiliation{Key Laboratory of Optical Astronomy, National Astronomical Observatories, Chinese Academy of Sciences, Beijing 100101, Peopleʼs Republic of China;}

\author[0009-0005-0716-0247]{Haiyang Mu}
\affiliation{National Astronomical Observatories, Chinese Academy of Sciences, Beijing 100101, People’s Republic of China}
\affiliation{Key Laboratory of Optical Astronomy, National Astronomical Observatories, Chinese Academy of Sciences, Beijing 100101, Peopleʼs Republic of China;}

 
\author[0000-0002-7397-811X]{Yunfei Xu}
\affiliation{National Astronomical Observatories, Chinese Academy of Sciences, Beijing 100101, People’s Republic of China}
\affiliation{National Astronomical Data Center, Beijing 100101, People’s Republic of China
}  

\author[0000-0002-8669-5370]{Dongwei Fan}
\affiliation{National Astronomical Observatories, Chinese Academy of Sciences, Beijing 100101, People’s Republic of China}
\affiliation{National Astronomical Data Center, Beijing 100101, People’s Republic of China
}  
   
\author{Guirong Xue}
\affiliation{Zhejiang Lab, Hangzhou 311100, People’s Republic of China}

\author{Ali Luo}
\affiliation{National Astronomical Observatories,
  Chinese Academy of Sciences,
  Beijing 100101, People’s Republic of China}

\author{Jifeng Liu}
\affiliation{Key Laboratory of Optical Astronomy, National Astronomical Observatories, Chinese Academy of Sciences, Beijing 100101, Peopleʼs Republic of China;}
\affiliation{Institute for Frontiers in Astronomy and Astrophysics, Beiing Normal University, Beijing 102206, People's Republic of China}
\affiliation{School of Astronomy and Space Science,
  University of Chinese Academy of Sciences,
  Beijing 100049, People’s Republic of China}
\affiliation{New Cornerstone Science Laboratory. NationalAstronomical Observatories, Chinese Academy of Sciences, Beiing 100012, People's Republic of China}

\footnotetext[1]{* Corresponding authors}



\begin{abstract}

Time-domain surveys have advanced astronomical research by revealing diverse variable phenomena, from stellar flares to transient events. The scale and complexity of survey data, along with the demand for rapid classification, present significant challenges for analysis. While machine learning offers solutions, most existing models are tailored to single tasks, struggle to generalize, and depend heavily on large, accurately labeled datasets. We introduce FALCO, a foundation model for astronomical light curve analysis in time-domain astronomy. This paper presents the primary version of FALCO trained via self-supervised learning on unlabeled Kepler light curves using a Transformer-based architecture. The model has been evaluated across three distinct light curve analysis tasks and it demonstrates robust performance in all tasks, achieving an accuracy of 95\% for stellar variability classification across eight classes, an overall RMSE of 0.1305 dex in surface gravity estimation (with significantly improved precision of RMSE $<$ 0.08 dex at low gravity end where log $g < 1$, and ~0.02 dex near log $g \approx 3$), and a precision of 87\% in flare identification. These results highlight the versatility of the foundation model in extracting generalizable representations from light curves, enabling easy adaptation to diverse tasks and making it a promising tool for time-domain analysis. Further analysis of model scaling and input light curve sequence length reveals that larger models and longer input sequences improve performance. We have also applied the model to produce a comprehensive catalog of surface gravity (log $g$) measurements for 179,732 Kepler stellar targets, using their light curves.


\end{abstract}

\section{Introduction} \label{sec:intro}



The emergence of modern time-domain surveys has enhanced our understanding of the dynamic universe, generating huge volumes of light curve data that encode rich information about transient and variable phenomena. These phenomena span a diverse range of astronomical objects, from stellar flares and exoplanets to gravitational wave electromagnetic counterparts \citep{aerts2010asteroseismology, yang2019flare, jin2020kilonova, cui2021identify, chen2023discovery, pan2024astroconformer, wang2024discovery}. The rapid advancement of astronomical survey missions, including Kepler, TESS, ZTF, and the Einstein Probe (EP) \citep{yuan2022einstein}, has initiated an era of data-intensive astronomy. This trend will intensify with the forthcoming operation of next-generation surveys, such as the Vera C. Rubin Observatory \citep{ivezic2019lsst} and the SiTian project \citep{liu2021sitian}, which will acquire light curves for an extensive population of celestial objects.

The synthesis of advanced observational tools and advanced data processing capabilities has enabled sky monitoring with enhanced temporal and spatial resolution. This technological evolution presents both opportunities and challenges for the astronomical community \citep{graham2012data, hlovzek2019data, coughlin2023data}. The diverse nature of astronomical phenomena and the exponential increase in observational data have driven the development of artificial intelligence(AI)-based methods as essential tools for efficient data analysis in time-domain astronomy.

Recent advancements in astronomical observation have highlighted the limitations of traditional machine learning methods, such as Convolutional Neural Networks (CNNs) and decision tree-based models. While these approaches have shown success in tasks like flare identification and variable star classification \citep{tu2022convolutional, zuo2024x, cui2024identifying}, they often rely on curated labeled datasets, necessitating significant manual effort and limiting their applicability to other astronomical phenomena. Furthermore, these models struggle with capturing long-range temporal dependencies and dealing with nonuniform sampling in astronomical time series.

In contrast, recent developments in AI, particularly Transformer-based architectures and large language models (LLMs), offer promising solutions for analyzing time series data \citep{donoso2023astromer, pan2024scaling, smith2024astropt}. The structural similarities between light curves and natural language sequences make them suitable for Transformer models, which can effectively capture complex temporal correlations with minimal labeled data through self-supervised learning.

Foundation models, initially designed for Natural Language Processing (NLP) \citep{devlin2018bert, radford2019language}, have shown remarkable generalization across various domains, including Computer Vision \citep{dosovitskiy2020image} and speech recognition \citep{baevski2020wav2vec}. Their self-supervised pre-training on large unlabeled datasets allows them to develop robust feature representations, which can be fine-tuned for specific tasks. This is particularly beneficial in astronomy, where labeled observations are scarce.

Recent studies have successfully adapted Transformer-based and foundation models for astronomy. Initial efforts focused on image analysis, such as AstroPT \citep{smith2024astropt}, while significant progress in time series analysis has emerged. For instance, ASTROMER \citep{donoso2023astromer} utilized Transformer architectures for light curve embeddings, and Astroconformer \citep{pan2024astroconformer} excelled in estimating stellar parameters. Additionally, research shows that these models can achieve greater sample efficiency compared to traditional methods \citep{pan2024scaling}, underscoring their potential in astronomical applications. \cite{pan2024scaling} investigates scaling laws in stellar light curves using self-supervised learning, demonstrating that a Transformer model offers markedly greater sample efficiency in inferring stellar surface gravity compared to traditional supervised learning approaches. 

Furthermore, multi-modal approaches like AstroM$^3$ \citep{rizhko2024astrom} integrate diverse data types, enhancing model performance. Domain-specific foundation models, such as Maven \citep{zhang2024maven} for supernova science, have demonstrated state-of-the-art results through innovative training methods. \cite{roy2024ai} introduced a foundation model for heliophysics, utilizing a Long-Short Spectral Transformer pretrained on Solar Dynamics Observatory data. These developments highlight the transformative potential of foundation models in advancing astronomical data analysis.

\begin{figure}[htbp]
    \centering
    \includegraphics[width=0.8\textwidth]{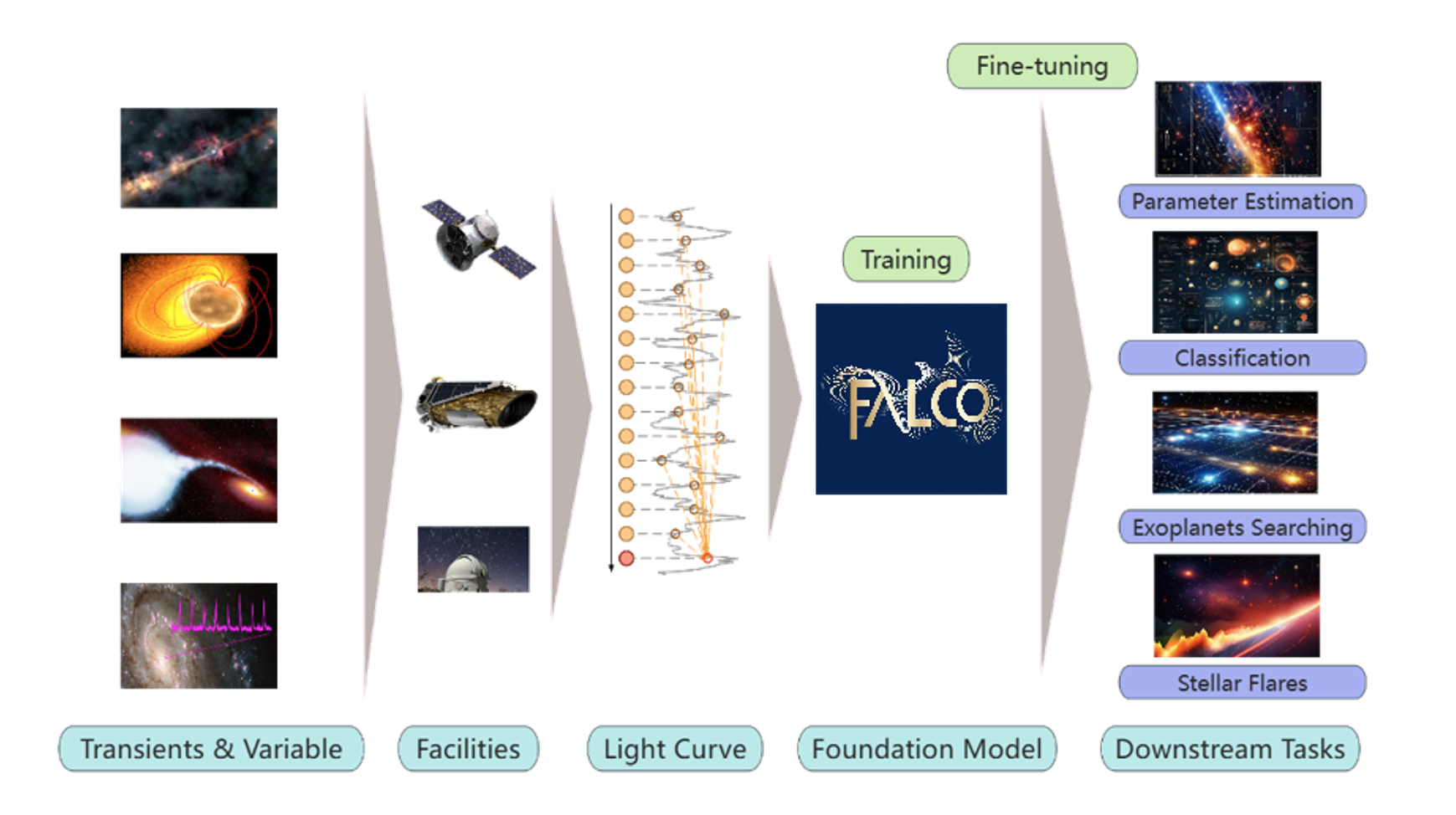} 
    \caption{The conceptual framework and roadmap of FALCO: From multi-source astronomical observations to specialized downstream science applications}

    \label{fig:overview}
\end{figure}

Motivated by the challenges of processing massive time-domain astronomical data and recent advances in Transformer-based models \citep{bommasani2021opportunities, pan2024scaling}, we propose FALCO, a foundation model specifically designed for light curve analysis in time-domain astronomy. Specifically, FALCO aims to address current challenges in three ways:
\begin{itemize}
 \item Employing a specialized transformer-based architecture optimized for processing complex time series data, effectively capturing both short-term variations and long-range dependencies in light curves.
\item Utilizing an advanced self-supervised learning strategy to maximize the use of unlabeled data, enabling robust feature learning from a variety of astronomical phenomena.
\item Providing a unified framework for various light curve analysis tasks, such as stellar variability classification, surface gravity estimation, and stellar flare identification.
\end{itemize}
These characteristics give FALCO advantages over existing approaches. As illustrated in Figure \ref{fig:overview}, FALCO can be applied in various light curve analysis tasks and excels in identifying diverse astronomical phenomena.

In this study, we implement FALCO using data from the Kepler mission \citep{borucki2010kepler}, validating the feasibility of constructing a foundation model for diverse light curve analysis tasks. The Kepler archive provides an ideal validation testbed for FAlCO due to its exceptional combination of high-precision photometric measurements, superior sensitivity and temporal resolution, long observation period and comprehensive sample size with diverse variability properties. Our model was evaluated across three distinct astronomical tasks, demonstrating outstanding performances in key metrics. This work establish FALCO as a promising tool for analyzing diverse time-domain sources in current and forthcoming large-scale surveys, such as TESS, ZTF, LSST and SiTian.

The structure of this paper is organized as follows. Section \ref{sec:data} introduces the data sources, preprocessing techniques, and datasets used in this study. Section \ref{sec:methodology} details the construction of the model and the methodologies employed for training it. Section \ref{scientific tasks} presents the results obtained from three scientific tasks. Section \ref{disscussion} offers a discussion and analysis of the findings. Section \ref{application} introduces the deployed applications and ideas of model applications. Section \ref{conclusion} concludes the study with final remarks and insights.

\section{Data} \label{sec:data}
This section details the data acquisition and processing methodology for both model training and scientific task evaluations. We utilize light curves from Kepler mission as an exemplary dataset for pre-training, and also constructing dataset for three typical light curve analysis tasks, namely stellar variability, surface gravity estimation, and flare detection. Since this study primarily focuses on validating approaches for developing a foundational model applicable to  various tasks, our model was trained and tested on Kepler data, without incorporating data from other surveys.

\subsection{Kepler Data}

The Kepler mission is a critical NASA initiative launched in 2009. Spanning from 2009 to 2013, the mission provided a dataset of light curves for nearly 200,000 stars at a consistent cadence of approximately 29.4 minutes. The Kepler mission successfully identified thousands of confirmed exoplanets, along with numerous additional planetary candidates, noticeably enhancing our understanding of planetary systems beyond our Solar System. It uncovered a diverse array of planetary architectures, including those markedly different from our own. Furthermore, data from Kepler have aided scientists in exploring the mechanisms of planet formation and evolution.

The operational lifespan of the Kepler mission is divided into two distinct phases. The initial phase, referred to as K1, involved the spacecraft's telescope being steadfastly directed towards a single star field in the Cygnus-Lyra region. Despite constraints related to data transmission efficiency, K1 successfully provided light curve data for an estimated 200,000 stars, covering up to 17 quarters.

Due to the uncontrolled behavior of its reaction wheels, the Kepler mission initiated its second phase, known as K2, in 2014, shifting its focus to observe stars around the ecliptic plane. This new phase introduced a novel observation paradigm, characterized by community-driven scientific objectives and shorter observational campaigns. However, data quality from K2 has been reported to be inferior to that of the initial phase (K1), as indicated by \cite{ilin2019flares}. The targets observed by K2 often exhibit a slow drift across the CCDs, followed by an abrupt return to their original positions, resulting in spikes in the light curves, as noted by \cite{van2016s}. This phenomenon complicates the analysis of short-term variations, particularly when conducting statistical studies of stellar flares. Consequently, our research relies on data from the first phase of the mission, K1, to ensure the reliability and consistency necessary for our analysis.

The Kepler light curves used in this study were retrieved from the Mikulski Archive for Space Telescopes (MAST)\footnote{\url{https://archive.stsci.edu/}}. We systematically processed the Kepler long-cadence quarterly data products, which consist of approximately 200,000 files in the Flexible Image Transport System (FITS) format, totaling around 94 GB. Each file includes headers that provide information about the observing conditions and data quality, along with tables of flux measurements over time. In this work, we employed the Pre-Search Data Conditioning Simple Aperture Photometry (PDC SAP) flux measurement, which effectively removes instrumental systematics and noise while preserving both stellar activity and exoplanet transit signals.

\subsection{Labeled Datasets}

This study employs three categories of labeled dataset for the downstream scientific tasks testing our model, namely stellar variability classification, estimation of surface gravity ($\log g$), and stellar flare identification. Table \ref{tab:light_curves_datasets} illustrate the number of light curves across datasets for various scientific tasks. Approximately half of the labeled data is used as unlabeled data, along with other unlabeled Kepler data, to train the FALCO model; the other half is used as labeled data for training and testing scientific tasks.

\begin{table}[htbp]
    \centering
    \caption{Number of Light Curves in Different Datasets}
    \label{tab:light_curves_datasets}
    \begin{tabular}{lllr}
        \toprule
        Dataset Type & Total Number of Light Curves & Available for Scientific Tasks & Reference \\
        \midrule
        Stellar Variability Classification  & 8,523 & 4,500 & \cite{audenaert2021tess} \\ 
        & & & \cite{paunzen2024apparent} \\ 
        Surface Gravity Estimation  & 17,750 & 8,875 & \cite{mathur2017revised} \\ 
        & & & \cite{yu2018asteroseismology} \\ 
        & & & \cite{yu2020asteroseismology} \\ 
        Stellar Flare Identification & 3,420 & 1,606 & \cite{yang2019flare} \\ 
        \bottomrule
    \end{tabular}
\end{table}

\subsubsection{Stellar Variability}\label{classification data}

In our study, the data and labels for the stellar variability classification task were derived from the work of \cite{paunzen2024apparent} and \cite{audenaert2021tess}. \cite{audenaert2021tess} presents a methodology for the automatic variability classification of TESS photometry using an ensemble of supervised learners combined into a metaclassifier. This approach was successfully validated with a meticulously curated labeled sample of Kepler Q9 light curves, which spanned a period of 27.4 days, simulating single-sector TESS observations. \cite{paunzen2024apparent} conducted a detailed analysis of non-variable stars from the Kepler mission, utilizing the Lomb-Scargle periodogram and false-alarm probability (FAP) to identify 14,154 non-variable stars. Since the “Constant” type in \cite{audenaert2021tess} is designed specifically for TESS, we instead incorporate the non-variable class introduced in \cite{paunzen2024apparent} as part of our dataset. Consequently, we used a portion of their non-variable star data to construct our stellar variability classification dataset.

\begin{figure}[ht]
    \centering
    \caption{Light curves of different astronomical classes derived from FITS files obtained via MAST from a single quarter of Kepler observations: APERIODIC, CONTACT ROT, DSCT BCEP, ECLIPSE, GDOR\_SPB, RRLYR CEPHEID, SOLARLIKE, and non-variable}
    \label{fig:different class}

    \begin{minipage}[t]{0.45\textwidth}
        \centering
        \includegraphics[width=\textwidth]{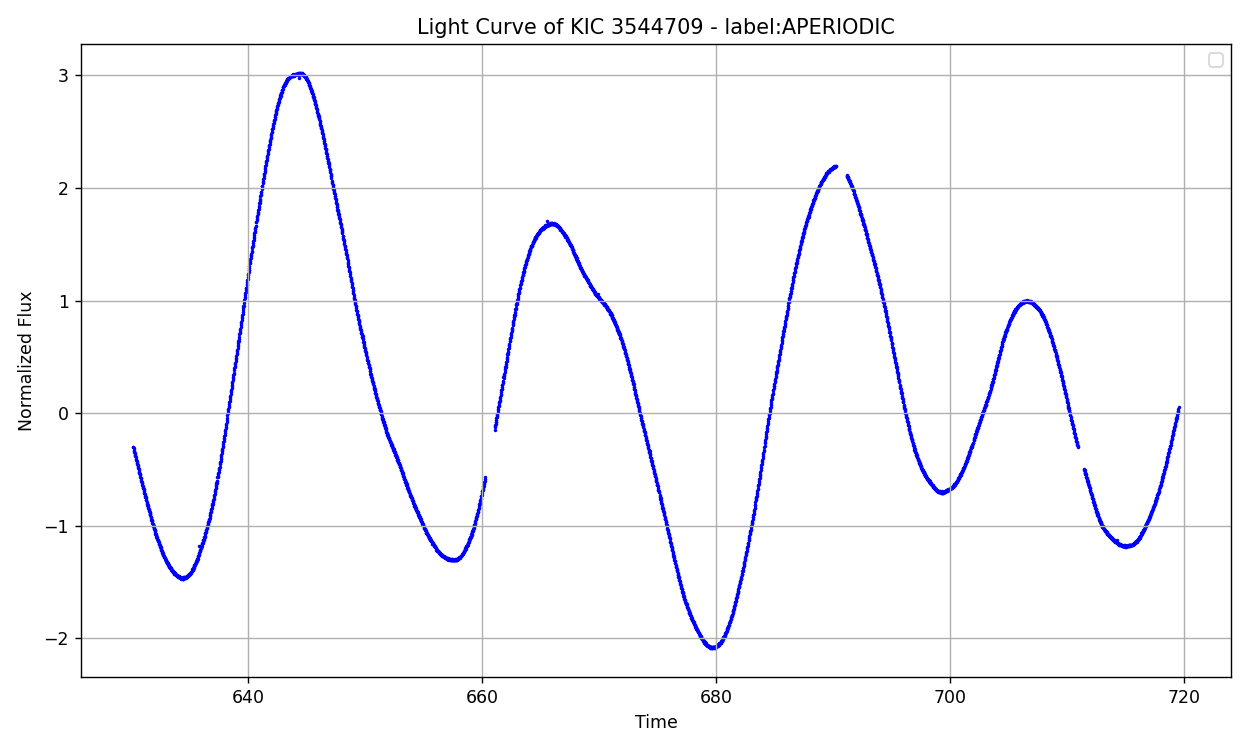}
        \vspace{0.3cm} 
        {\footnotesize (a) Light curve of Aperiodic: Characterized by irregular fluctuations without a clear periodic pattern.} 
        \label{fig:APERIODIC}
    \end{minipage}
    \hfill
    \begin{minipage}[t]{0.45\textwidth}
        \centering
        \includegraphics[width=\textwidth]{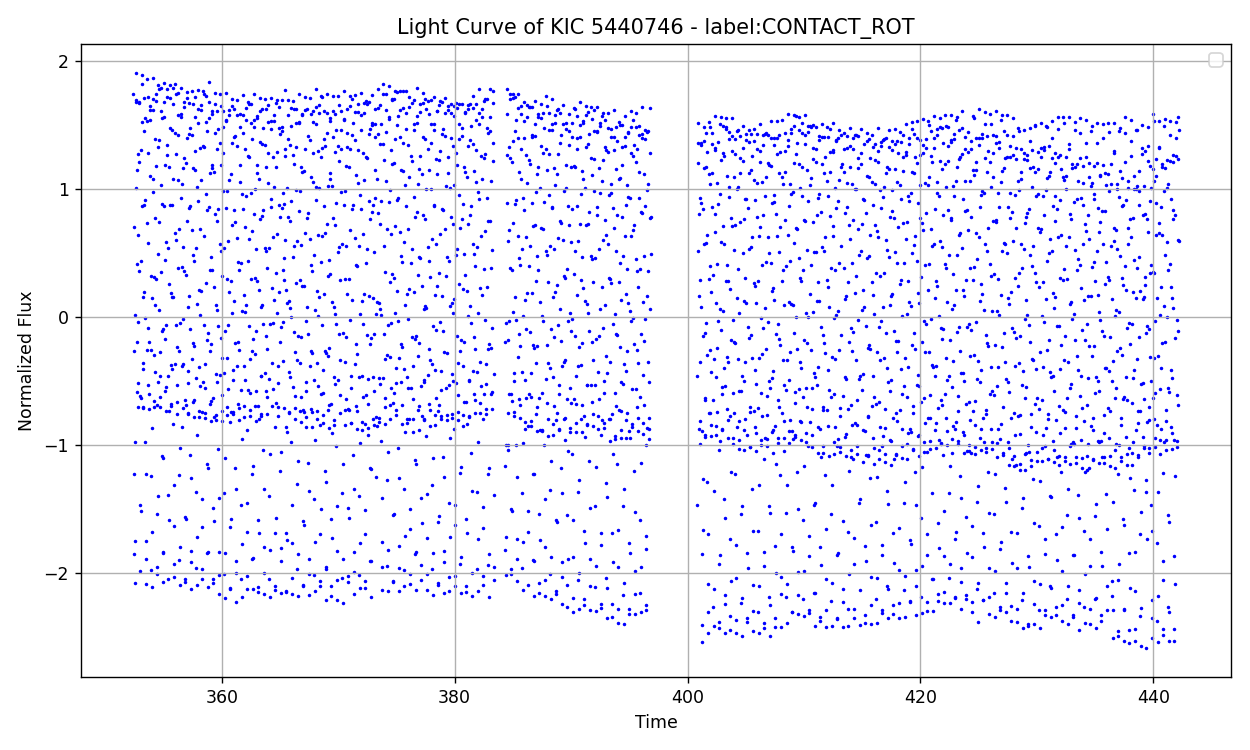}
        \vspace{0.3cm} 
        {\footnotesize (b) Light curve of CONTACT\_ROT: Exhibits a noisy, non-periodic pattern.}
        \label{fig:CONTACT_ROT}
    \end{minipage}

    \medskip 

    \begin{minipage}[t]{0.45\textwidth}
        \centering
        \includegraphics[width=\textwidth]{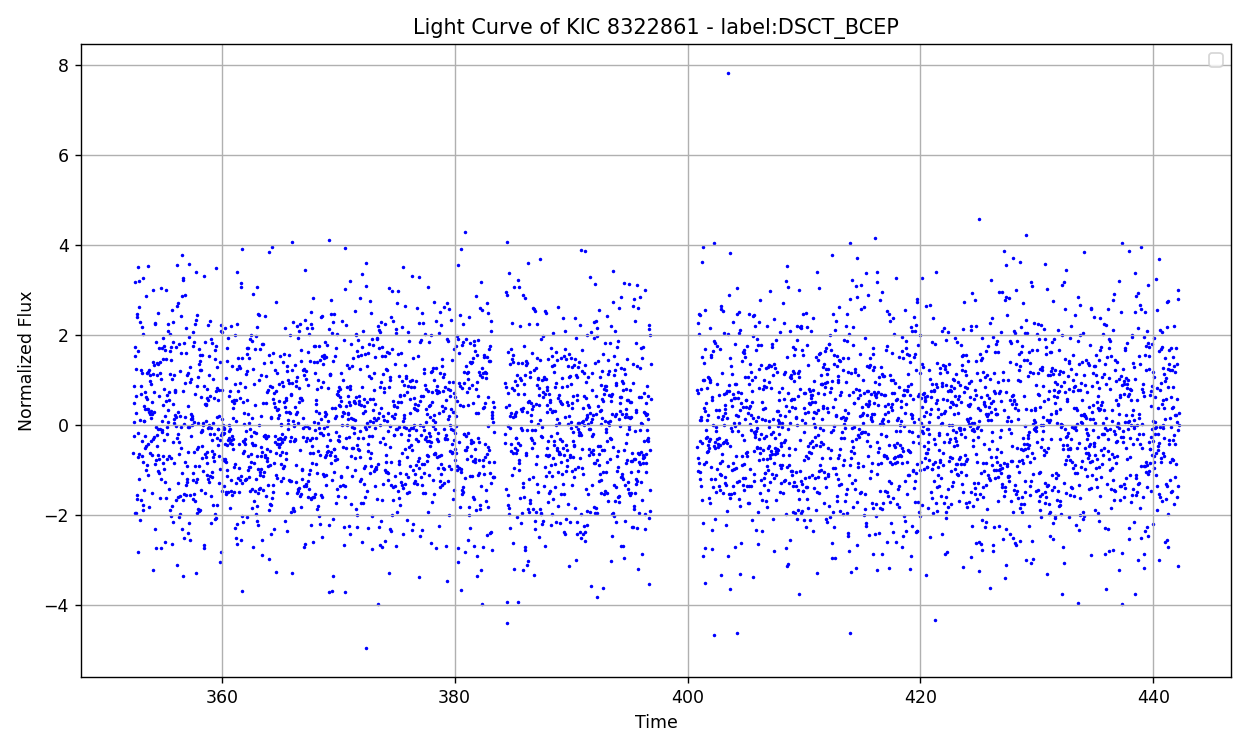}
        \vspace{0.3cm} 
        {\footnotesize (c) Light curve of DSCT\_BCEP: Displays a high-frequency oscillation pattern.}
        \label{fig:DSCT_BCEP}
    \end{minipage}
    \hfill
    \begin{minipage}[t]{0.45\textwidth}
        \centering
        \includegraphics[width=\textwidth]{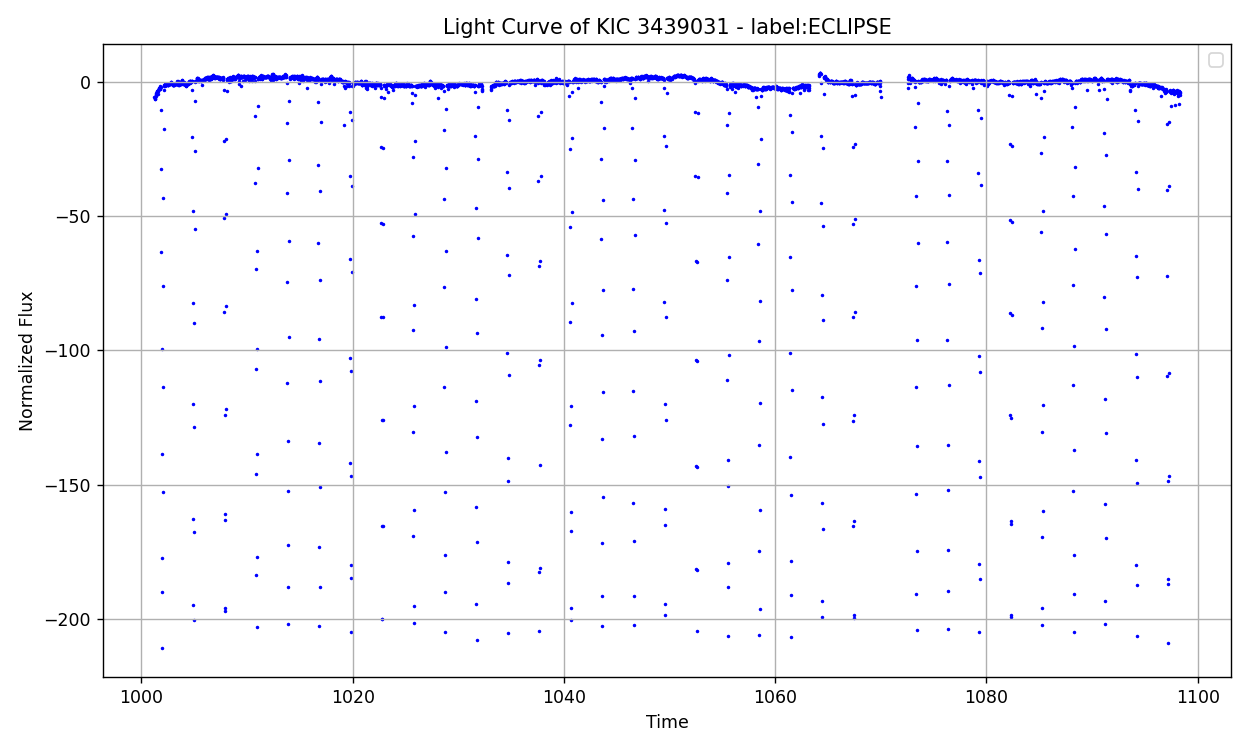}
        \vspace{0.3cm} 
        {\footnotesize (d) Light curve of ECLIPSE: Shows periodic dips in flux.}
        \label{fig:ECLIPSE}
    \end{minipage}

    \medskip 

    
    \begin{minipage}[t]{0.45\textwidth}
        \centering
        \includegraphics[width=\textwidth]{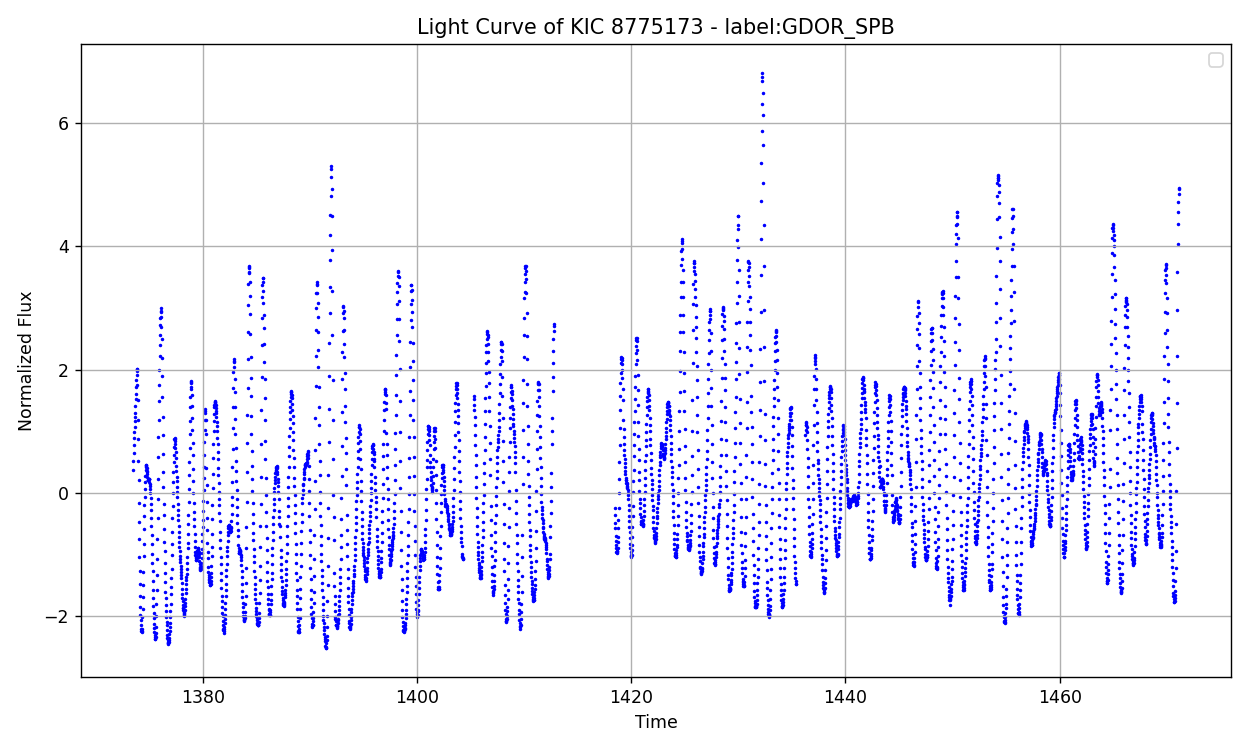}
        \vspace{0.3cm} 
        {\footnotesize (f) Light curve of GDOR\_SPB: Demonstrates complex, multi-periodic variations.}
        \label{fig:GDOR_SPB}
    \end{minipage}
    \hfill
    \begin{minipage}[t]{0.45\textwidth}
        \centering
        \includegraphics[width=\textwidth]{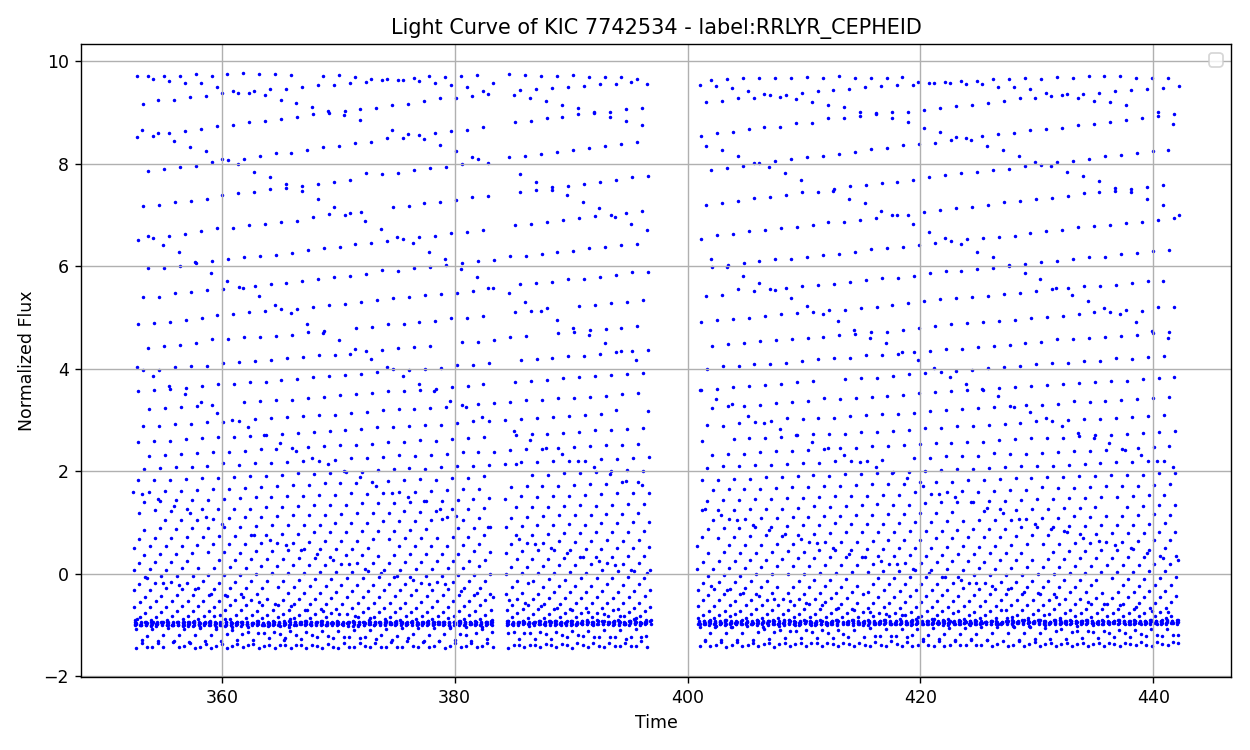}
        \vspace{0.3cm} 
        {\footnotesize (h) Light curve of RRLYR\_CEPHEID: Exhibits a distinctive sawtooth pattern.}
        \label{fig:RRLYR_CEPHEID}
    \end{minipage}

    \medskip 


    \begin{minipage}[t]{0.45\textwidth}
        \centering
        \includegraphics[width=\textwidth]{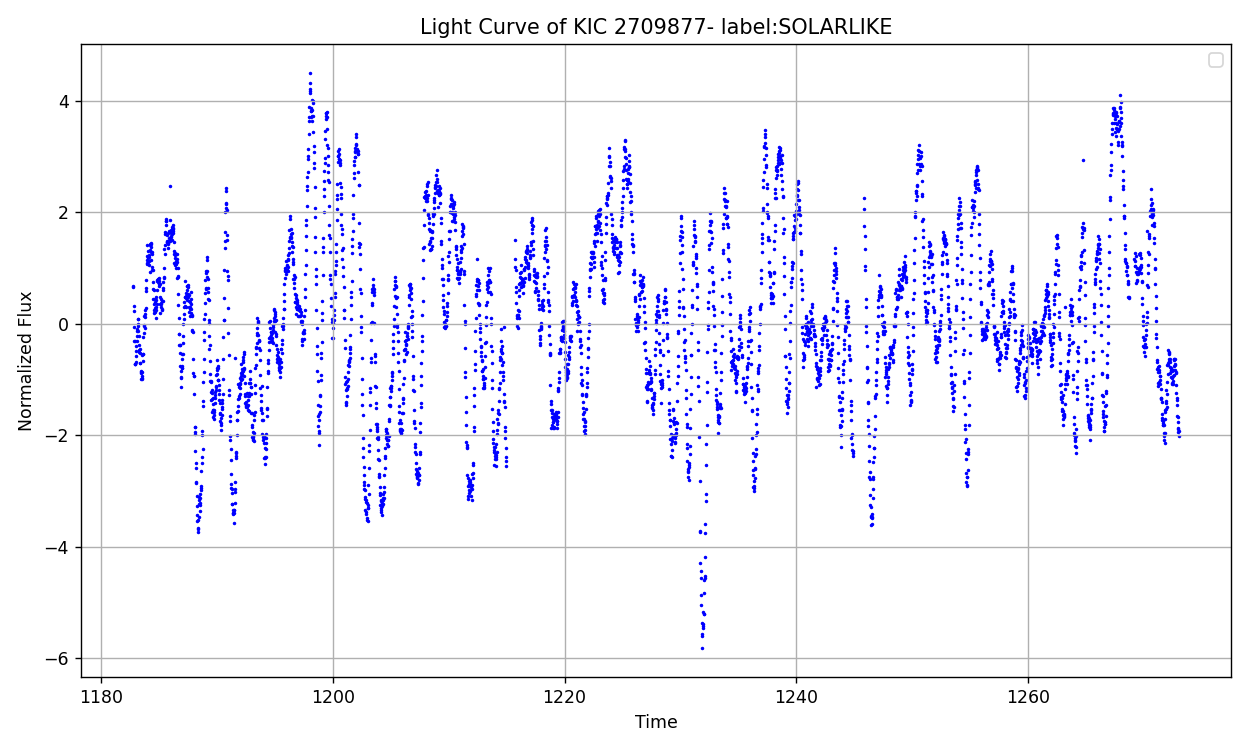}
        \vspace{0.3cm} 
        {\footnotesize (e) Light curve of SOLARLIKE: Features low-amplitude oscillations.}
        \label{fig:solarlike}
    \end{minipage}
    \hfill
    \begin{minipage}[t]{0.45\textwidth}
        \centering
        \includegraphics[width=\textwidth]{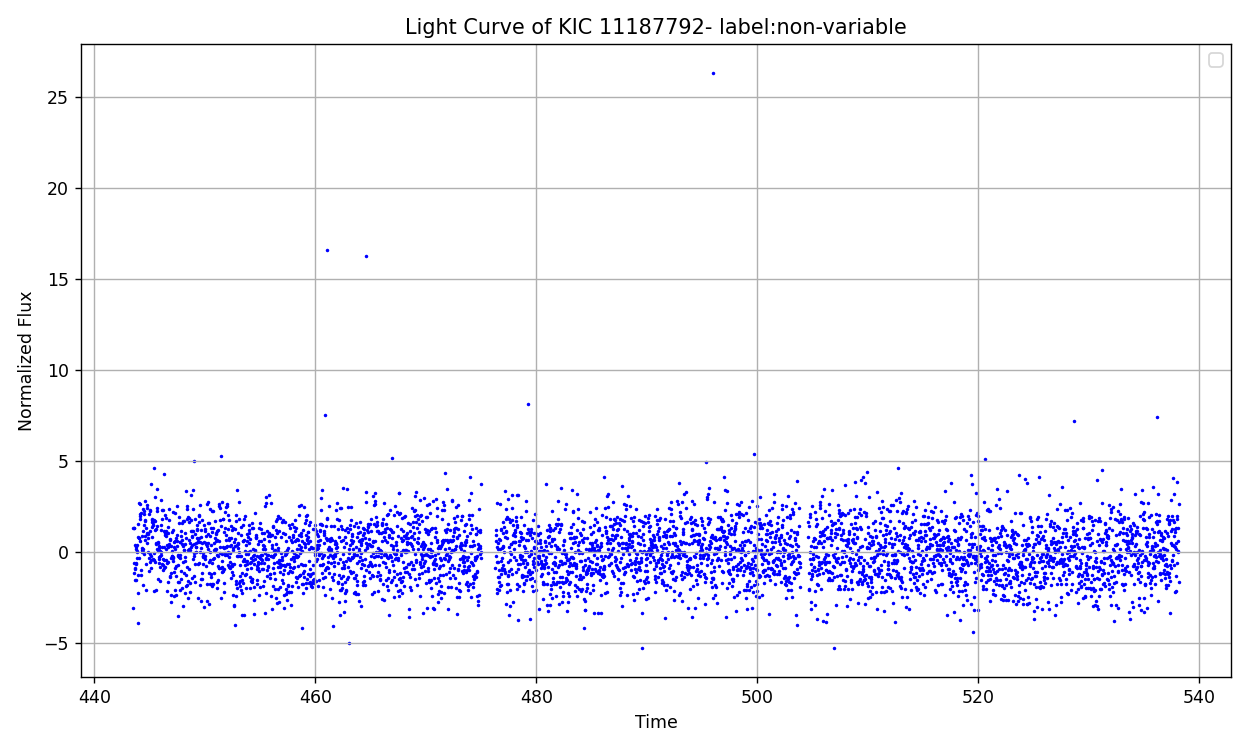}
        \vspace{0.3cm} 
        {\footnotesize (g) Light curve of non-variable: Represents a stable light curve with minimal fluctuation.}
        \label{fig:non-variable}
    \end{minipage}
\end{figure}

This dataset encompasses eight distinct classes of celestial objects from \cite{paunzen2024apparent} and \cite{audenaert2021tess}. The classifications include: Aperiodic stars (APERIODIC), Contact binaries and rotational variables (CONTACT ROT), $\delta$ Scuti and $\beta$ Cephei stars (DSCT BCEP), Eclipsing binaries (ECLIPSE), $\gamma$ Doradus and Slowly Pulsating B-type stars (GDOR\_SPB), RR Lyrae and Cepheid variables (RRLYR CEPHEID), Solar-like pulsators (SOLARLIKE), and Non-variable Class (non-variable). Detailed statistics regarding the number and proportion of objects within each category are summarized in Table \ref{tab:training_set}. Figure \ref{fig:different class} displays representative light curves of each class, with temporal evolution plotted on the x-axis and normalized flux on the y-axis.

\begin{table}[htbp]
\centering
\caption{Description of each label, type, number, and proportion in the stellar variability classification dataset.}
\label{tab:training_set}
\begin{tabular}{llrr}
\toprule
\textbf{Class label} & \textbf{Type} & \textbf{Number} & \textbf{Proportion (\%)} \\
\midrule
APERIODIC & Aperiodic stars & 830 & 9.73 \\
CONTACT ROT & Contact binaries and rotational variables & 2260 & 26.52 \\
DSCT BCEP & $\delta$ Sct and $\beta$ Cep stars & 772 & 9.06 \\
ECLIPSE & Eclipsing binaries & 974 & 11.43 \\
GDOR\_SPB & $\gamma$ Doradus and SPB stars & 630 & 7.39 \\
RRLYR CEPHEID & RR Lyraes and Cepheids & 62 & 0.73 \\
SOLARLIKE & Solar-like pulsators & 1800 & 21.12 \\
non-variable & Non-Variable stars & 1195 & 14.04 \\
\midrule
\textbf{Total} & \textbf{-} & \textbf{8,523} & \textbf{100} \\
\bottomrule
\end{tabular}
\end{table}

\subsubsection{Surface Gravity}\label{logg data}

Inspired by the work of \cite{pan2024scaling}, we also evaluate the performance of our model on the $\log g$ estimation task. The labeled data originated from the same source as referenced in \cite{pan2024scaling}, specifically incorporating asteroseismic measurements from \cite{sayeed2021swan} and \cite{yu2018asteroseismology, yu2020asteroseismology}.

\cite{yu2018asteroseismology} conducted a systematic characterization of solar-like oscillations and granulation for 16,094 oscillating red giants, utilizing end-of-mission long-cadence data. The stars in their study exhibited $\log g$ values ranging from 1.5 to 3.3. The typical uncertainties associated with their estimations of $\nu_{\text{max}}$ and asteroseismic $\log g$ were 1.6\% and 0.01–0.02 dex, respectively.

Subsequently, \cite{sayeed2021swan} performed a cross-referencing of two asteroseismic surface gravity catalogs: one from \cite{mathur2017revised} and the other from \cite{yu2018asteroseismology}, to enrich the dataset with detailed and corroborated $\log g$ values.

\begin{figure}[htbp]
    \centering
    \includegraphics[width=0.5\textwidth]{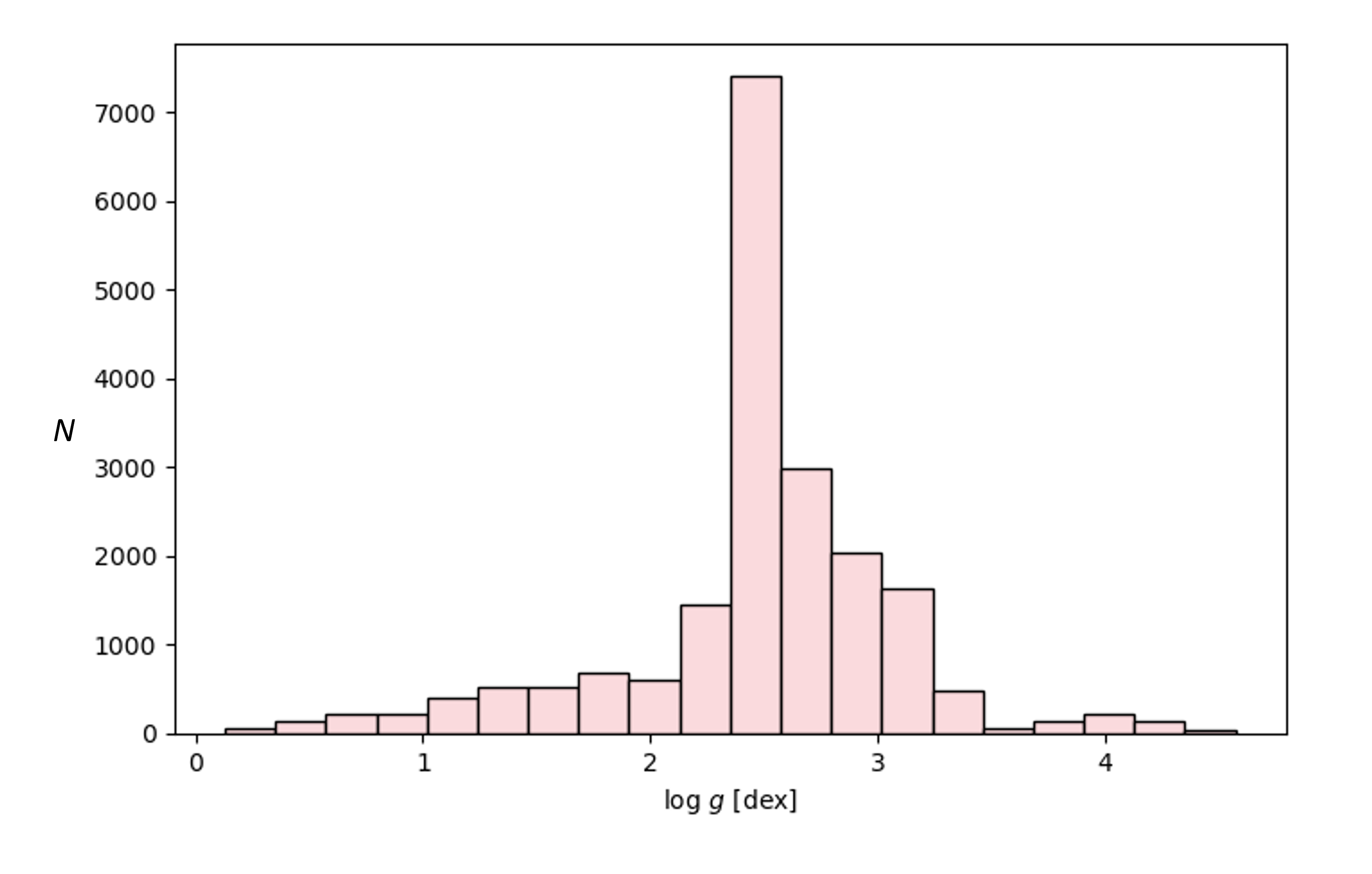} 
    \caption{The histogram of the distribution of log $g$ values within the surface gravity dataset.}
    \label{fig:loghist}
\end{figure}

Ultimately, we have curated a dataset comprising 17,750 high-quality light curves. The histogram depicted in Figure \ref{fig:loghist} represents the distribution of $\log g$ values within the dataset. We allocated half of this dataset for pre-training and the other half for downstream scientific task evaluation. The distribution of $\log g$ values across these light curves is illustrated in Figure \ref{fig:loghist}, providing a visual representation of the data's diversity and quality.

\subsubsection{Stellar Flare}\label{flare data}

The task of flare identification employs the systematically validated catalog from \cite{yang2019flare}. This dataset, derived from the long-cadence observations of Kepler DR25, encompasses 3,420 flare stars and documents 162,262 flare events. The identification of flare candidates is based on the criterion that after detrending baselines, at least three consecutive points exceeding 3$\sigma$ with no break points within 3 hours are required, concluding at the last point above 1$\sigma$. Previous research on flare identification predominantly adopted a three-step methodology: fitting the quiescent flux or establishing a baseline to mitigate noise and artifacts, applying threshold criteria for flare identification, and validating the results while checking for potential sources of contamination \citep{hawley2014kepler,gao2016white,yang2017flaring}. \cite{yang2019flare} introduced an enhanced method for flare identification that effectively reduces false-positive signals and instrumental errors, offering a more detailed and accurate flare catalog compared to its predecessors. Consequently, we selected this refined flare dataset from the complete Kepler data archive for subsequent scientific analysis.

\begin{figure}[ht]
    \centering
    \begin{minipage}{0.32\textwidth}
        \includegraphics[width=\textwidth]{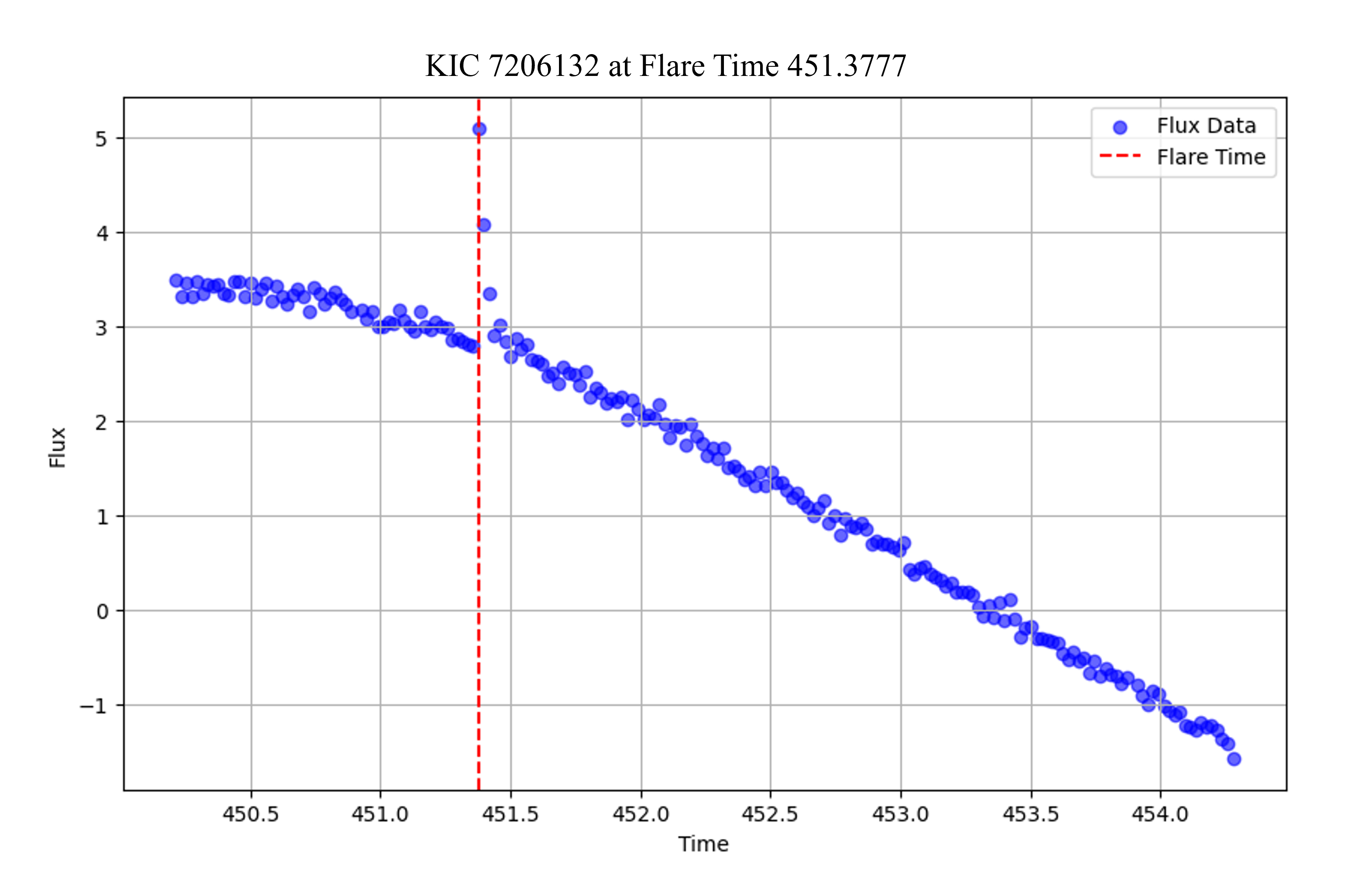}
        \label{fig:flare1}
    \end{minipage}
    \hfill
    \begin{minipage}{0.32\textwidth}
        \includegraphics[width=\textwidth]{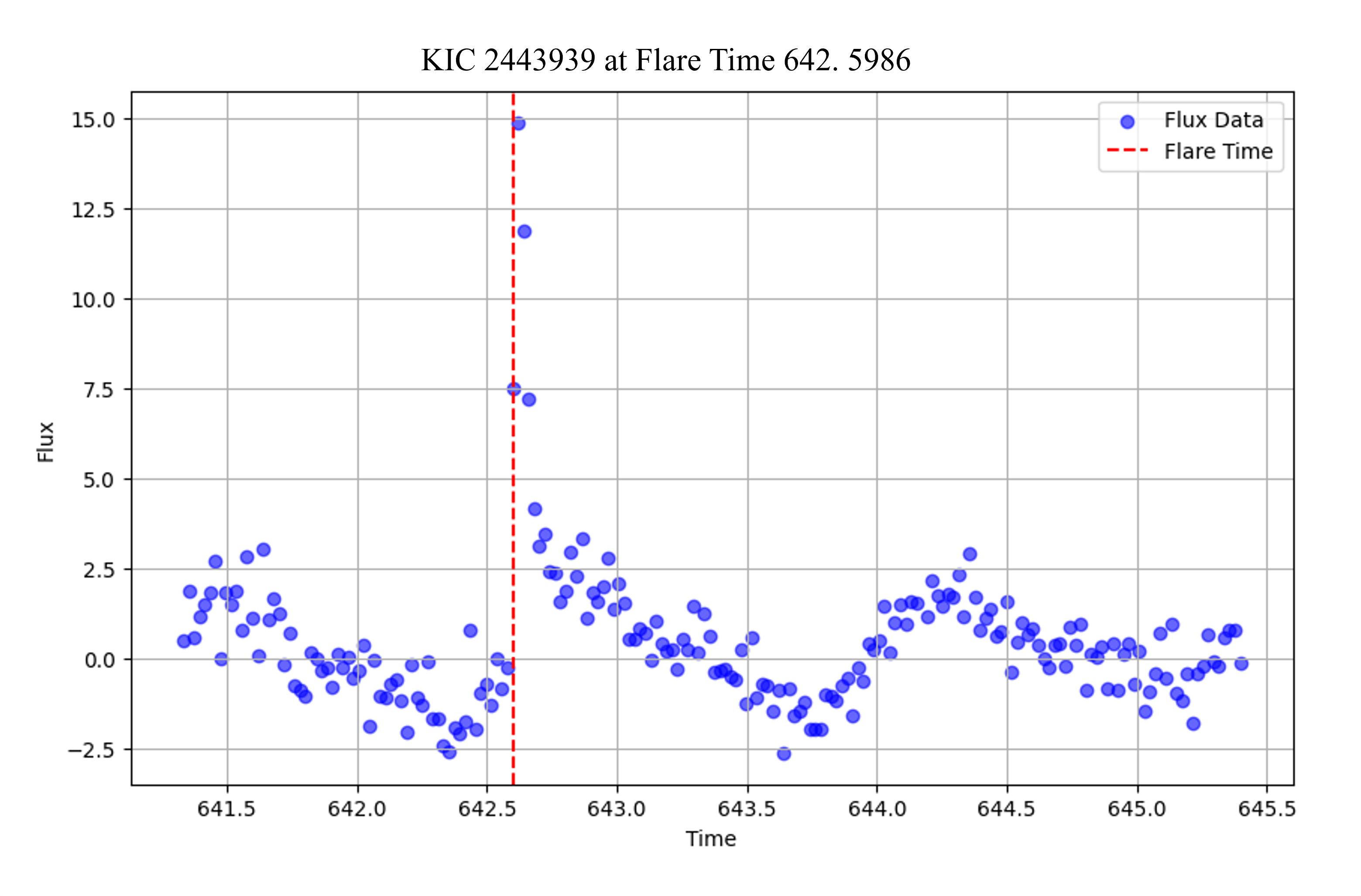}
        \label{fig:flare2}
    \end{minipage}
    \hfill
    \begin{minipage}{0.32\textwidth}
        \includegraphics[width=\textwidth]{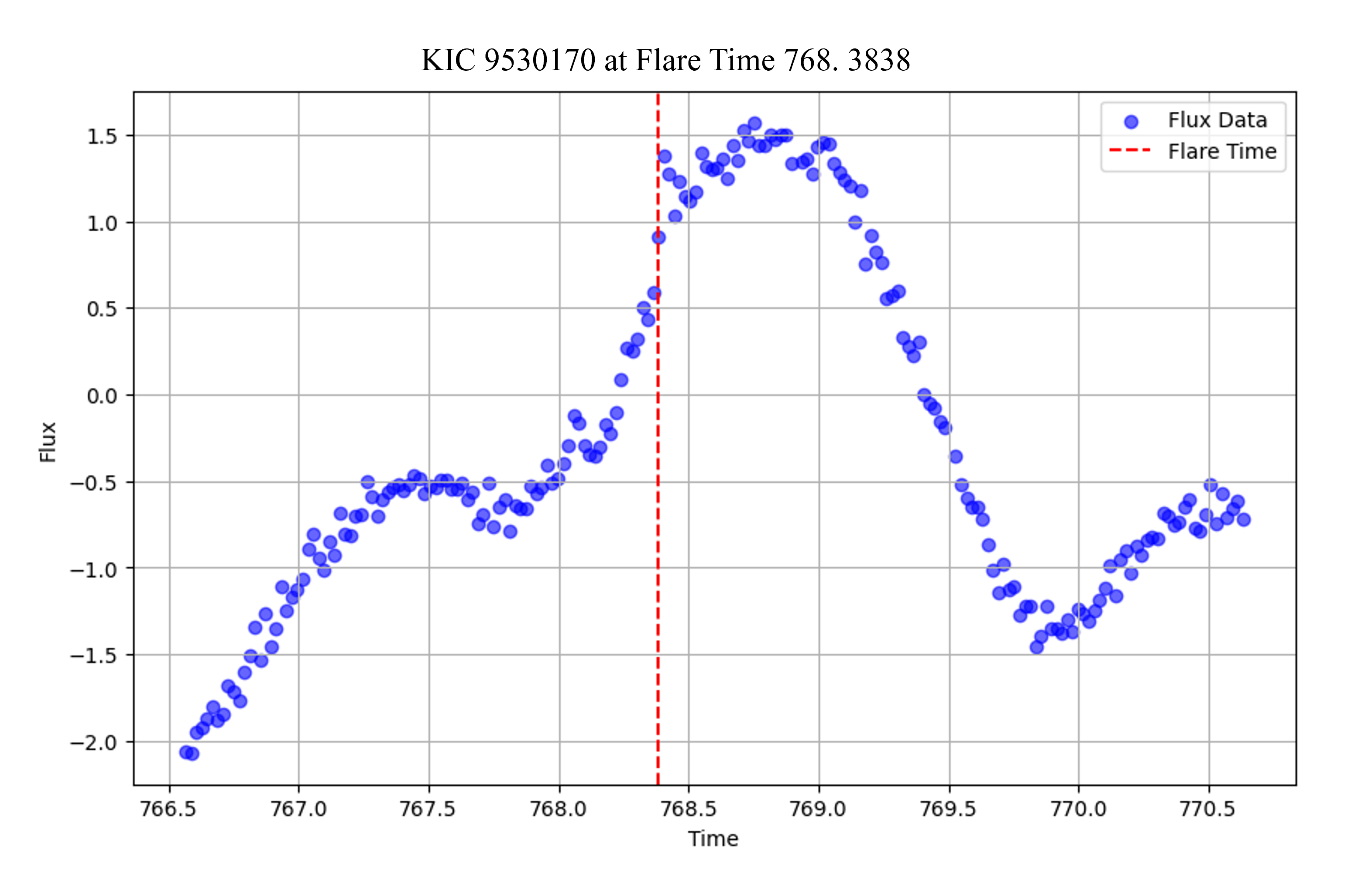}
        \label{fig:flare3}
    \end{minipage}
    \caption{Examples of light curves for flare events with a length of 200 points; labels from \cite{yang2019flare}}
    \label{fig:flare_data}
\end{figure}

In the flare identification part, the objective is to determine whether a given light curve segment exhibits a flare event. Since flares represent short-term, instantaneous phenomena, the dataset handling for such events differs somewhat from other scientific tasks. Instead of utilizing the entire light curve, we focused on segments where flare occurrences were identified. The input segments are standardized to a length of 200 data points. For this binary classification task, we label segments from Kepler light curves that have never experienced a flare as class 0, while segments that include at least one flare event are labeled as class 1. Given that flares are sporadic events, the dataset for flare identification is highly imbalanced, with instances labeled as 1 (flare events) constituting only a small fraction of the total dataset. Table \ref{tab:flaredataset_distribution} provides the distribution of flare and non-flare events across the training and test sets, along with the percentage representation of each class. 

\begin{table}[htbp]
    \centering
    \caption{Numbers of Flare and No Flare Events in stellar flare dataset}
    \begin{tabular}{lrrr}
        \toprule
        Class & Number of Train set & Number of Test set & Percentage\\
        \midrule
        No Flare (0) &  1,672,940 & 418,236 & 97.16\% \\
        Flare (1)&  48,878  & 12,219  & 2.84\% \\
        \bottomrule
    \end{tabular}
    
    \label{tab:flaredataset_distribution}
\end{table}

\subsection{Pre-training Datasets} \label{pre-train data}

We implemented a rigorous data partitioning strategy to evaluate model generalizability across scientific tasks. For the stellar variability classification, surface gravity (log $g$) estimation, and stellar flare identification, 50\% of each dataset was allocated to a hold-out test set strictly isolated from training procedures, as shown in the third column of Table \ref{tab:light_curves_datasets}.
This approach guarantees that the model is evaluated on completely independent data, enhancing its generalization capability. During the pre-training phase, a total of 165,462 light curves were utilized. After randomly shuffling the dataset, 80\% of the light curves were allocated to the training set, resulting in 132,369 light curves for training purposes, while the remaining 20\% comprised 33,093 light curves designated for the validation set.

\subsection{Light Curve Preprocessing}

In contrast to the detailed preprocessing implemented by \cite{pan2024scaling}, which was specifically tailored to log $g$ estimation task through sigma-clipping, Savitzky–Golay filtering \citep{schafer2011savitzky}, and 30-minute interval data interpolation, in this work we adopt a minimalist approach to data preparation.  This minimalist strategy aims to evaluate the effectiveness of raw data in training foundation models, challenging the conventional reliance on complex preprocessing techniques.

By simplifying the preprocessing phase, we seek to determine if a reduction in complexity can still yield robust model performance. This approach not only facilitates a more efficient workflow but also contributes to the development of more generalizable methodologies in light curve analysis. The key preprocessing steps for preparing high-quality light curve data from the Kepler mission are as follows:

\begin{enumerate}
    \item \textbf{Extraction of PDC SAP Flux and Time:} We extracted the PDC SAP flux and time series data from the full-volume FITS files provided by the Kepler mission.
    \item \textbf{Normalization and Concatenation Across Quarters:} Since Kepler observations are divided into multiple quarters, with variations in data quality due to instrumental effects, we applied Median Absolute Deviation (MAD) normalization to the data from each quarter separately before concatenation. This ensured a consistent scale across quarters, yielding complete light curves for each source. The MAD normalization formula is given by:
    \begin{equation}
    flux' = \frac{flux - \text{median}(flux)}{\text{MAD}(flux)}
    \end{equation}
    
    where:
    \begin{itemize}
        \item $flux$ represents the original flux data points,
        \item $\text{median}(flux)$ is the median of the flux,
        \item $\text{MAD}(flux)$ is the Median Absolute Deviation, defined as:
        \begin{equation}
            \text{MAD}(flux) = \text{median}\left( |flux - \text{median}(flux)| \right)
        \end{equation}
    \end{itemize}

    \item \textbf{Quality Control and Data Length Screening:} To ensure high data quality and sufficient length of light curves, we filtered out those with fewer than approximately 125,000 data points. As a result, we retained a total of 179,679 high-quality light curves.
    \item \textbf{Segmentation Based on Quarter Boundaries:} Due to significant gaps between observation quarters, we segmented the concatenated light curves based on the start and end times of each quarter to maintain temporal integrity.
    \item \textbf{Handling Missing Values:} Each light curve was inspected for NaN values, which were then imputed using linear interpolation to maintain the continuity of the time series.
    \item \textbf{Conversion to Torch Tensors and Data Partitioning:} Finally, all light curve data were converted into \textit{torch.tensor} format and partitioned into segments of varying lengths (e.g., 200, 100, 50 and 30 data points) to accommodate the requirements of different training scenarios.

\end{enumerate}

\section{Methodology} \label{sec:methodology}


As a foundation model designed for learning generic representations applicable to multiple astronomical time-series analysis tasks, FALCO employs self-supervised pre-training on extensive unlabelled light curve data. The model generates embeddings that capture intricate temporal patterns without explicit supervision, establishing a framework for diverse downstream applications.


The pre-trained foundation model can adapt to downstream scientific tasks through different transfer learning paradigms. In this work, we employed the model by extracting features from intermediate layers and applying task-specific MLPs to address different tasks. By extracting features from intermediate layers and applying task-specific MLPs, we systematically evaluated the model across different downstream astronomical tasks. This approach not only validates the foundation model paradigm in astronomical applications but also offers practical guidelines for leveraging pre-trained models in future astronomical research.

\subsection{Model Architecture} \label{model}

The Transformer-based architecture adopted in this study is inspired by the work of \cite{pan2024scaling}, which adapts the GPT-2 architecture for time series modeling. While the original implementation is based on nanoGPT\footnote{https://github.com/karpathy/nanoGPT}, the architecture is modified to handle continuous light curve data instead of discrete tokens.

The model architecture comprises a stack of $N$ Transformer blocks, each designed with multi-head self-attention mechanisms, position-wise feed-forward networks, layer normalization layers, and residual connections. To tailor this architecture for light curve analysis, we implement key modifications in the input processing stage. Specifically, token embeddings are replaced with a continuous time series encoder, and positional embeddings are adapted to effectively capture temporal relationships. The attention patterns within the blocks are specifically modified to reflect the characteristics of time series data. For input and output processing, we utilize two specialized MLPs: the first maps raw light curve data to the model dimension while preserving temporal information and accommodating variable-length sequences; the second transforms the features into task-specific outputs, supporting continuous value prediction while ensuring physical interpretability.


\begin{figure}[htbp]
    \centering
    \includegraphics[width=0.8\textwidth]{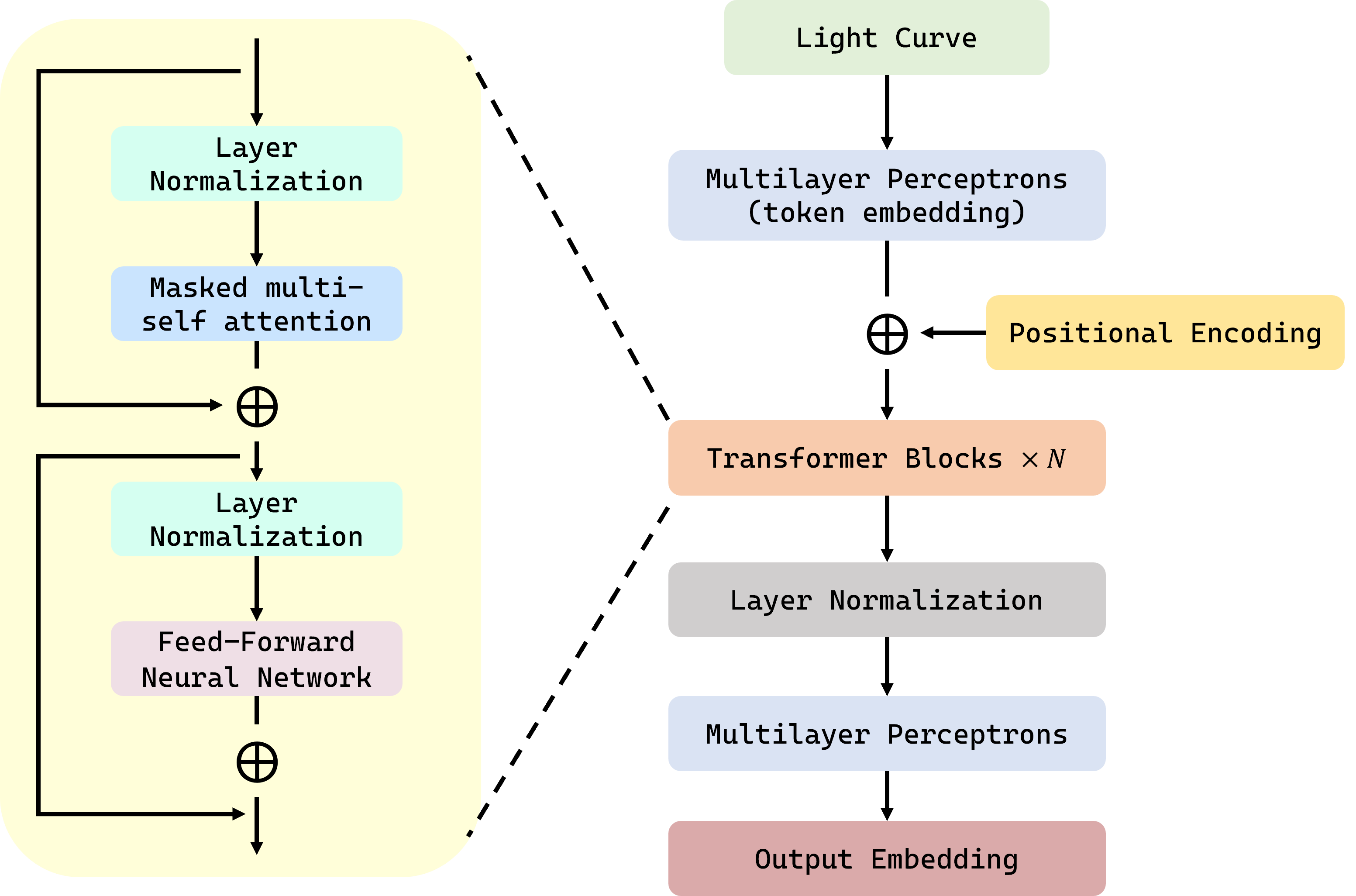} 
    \caption{The architecture of this model, the $N$ in the figure represents the number of Transformer blocks}
    \label{fig:architecture}
\end{figure}

Our adapted architecture, as illustrated in Figure \ref{fig:architecture}, fundamentally differs from traditional language models in several aspects. Instead of processing discrete tokens, our model handles continuous values throughout its pipeline. The attention mechanisms are specifically tailored for time series analysis, complemented by specialized input/output transformations. These architectural adaptations enable effective processing of astronomical time series data while leveraging the powerful pattern recognition capabilities inherent in Transformer architectures.

\subsection{Pre-training Settings}

The model was trained using the dataset described in Section \ref{pre-train data}. Following common practice in Transformer training, the learning rate was adjusted using cosine annealing with linear warm-up \citep{loshchilov2016sgdr}.The optimization process employed the AdamW optimizer \citep{loshchilov2017decoupled} with gradient clipping at 1.0, complemented by early stopping criteria. The complete training configuration, including hyperparameters, is detailed in Table \ref{tab:training-parameters}.

Given that categorical cross-entropy loss is unsuitable for the continuous nature of light curve data, this work followed \cite{pan2024scaling} in adopting Huber loss \citep{huber1992robust}. The Huber loss is defined as:  
\begin{equation}
    L_\delta(y, f(x)) =
    \begin{cases} 
        \frac{1}{2}(y - f(x))^2 & \text{for } |y - f(x)| \leq \delta, \\
        \delta \left( |y - f(x)| - \frac{1}{2}\delta \right) & \text{otherwise.}
    \end{cases}
    \label{eq:huber_loss}
\end{equation}
where $y$ is the true value, $f(x)$ is the predicted value, and $\delta$ is a hyperparameter that determines the point at which the loss transitions from quadratic to linear. This loss function combines the advantages of mean squared error (MSE) and mean absolute error (MAE): it uses the squared loss (similar to MSE) for small errors ($|y - f(x)| \leq \delta $), promoting faster convergence with smooth gradients, while transitioning to a linear loss (similar to MAE) for larger errors, providing robustness against outliers. This formulation effectively models the conditional distribution under a constant-variance Gaussian assumption, making it particularly suitable for continuous time series prediction.

\begin{table}[htbp]
\centering
\caption{Hyperparameter settings for model training.}
\begin{tabular}{lll}
\toprule
\textbf{Parameter Name} & \textbf{Parameter Value} & \textbf{Description} \\
\midrule
Learning Rate Decay        & True                     & Whether to enable learning rate decay \\
Initial Learning Rate      & $2 \times 10^{-4}$     & Learning rate at the end of the warm-up phase \\
Warm-up Iterations         & 2,000                    & Number of iterations for linearly increasing the learning rate \\
Cosine Decay Iterations    & 60,000                   & Number of iterations for cosine decay of the learning rate \\
Minimum Learning Rate      & $2 \times 10^{-5}$     & Minimum value of the learning rate \\
Total Iterations           & 500,000                  & Total number of training iterations \\
Early Stopping Condition 1 & 2,000                    & Stop training if validation loss does not improve for 2,000 iterations \\
Early Stopping Condition 2 & 250,000                  & Stop training if validation loss exceeds half of the total iterations \\
Optimizer Type             & AdamW                    & Optimizer used for parameter updates \\
$\beta_1$                & 0.9                      & First momentum coefficient of the optimizer \\
$\beta_2$                & 0.95                     & Second momentum coefficient of the optimizer \\
Weight Decay               & 0.1                      & Weight decay coefficient for regularization \\
Maximum Gradient Norm      & 1.0                      & Maximum norm for gradient clipping \\
$\delta$ in Huber Loss             & 1.0            & The point at which the loss transitions from quadratic to linear \\
Single Pass                & True                     & Model trains on all data in a single pass \\
\bottomrule
\end{tabular}

\label{tab:training-parameters}
\end{table}


The model was trained for one epoch, iterating once through all training samples with a batch size of 32. With point-wise tokenization, this generates 7.6 billion tokens in total - 6.1 billion for training (132,369 light curves) and 1.5 billion for validation (33,093 light curves).


We explored the effects of varying input lengths on model performance. By training our models on light curves with different input lengths, we aimed to assess how these variations influenced the outcomes across various scientific tasks. Through comparative studies using input lengths of 30, 50, 100, and 200 time steps, several lengths were analyzed to assess their impact on model performance in capturing information and patterns within light curves, with the complete analysis presented in Section \ref{length}.

\begin{figure}[htbp]
    \centering
    \includegraphics[width=0.7\textwidth]{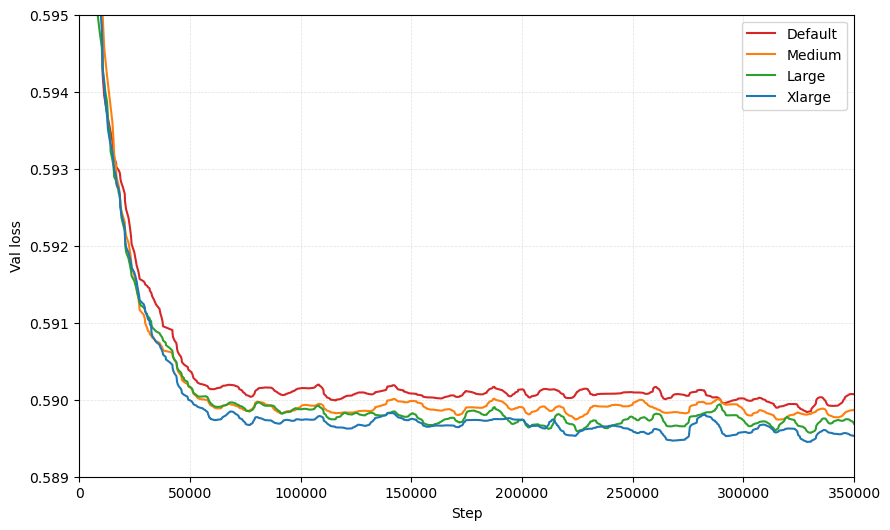} 
    \caption{During the model training process, we expect both the training loss and the validation loss to decrease. Across several training runs, both losses initially declined precipitously, then gradually leveled off around 50,000 iterations, after which they continued to decrease at a much slower rate. It illustrates the trend of the validation loss reduction and compares the efficacy of loss decrease among models with different parameter scales.}
    \label{fig:val loss}
\end{figure}

We also analyze the model scaling of light curve modeling by training models of varying sizes. The architectural specifications follow the model configurations established in \cite{radford2019language}, with different scales achieved by varying three fundamental parameters: number of Transformer blocks (depth), hidden layer dimension (width), and number of attention heads per block, as detailed in Table \ref{tab:model_params}. These parameters directly determine the model's capacity and computational requirements. Figure \ref{fig:val loss} illustrates the trend of the validation loss reduction and compares the efficacy of loss decrease among models with different parameter scales. The analysis of model scaling is detailed in Section \ref{sclaing law}.

The results indicated that larger models and longer input lengths of light curves generally achieved better performance. This exploration provided valuable insights into the interplay between model complexity and input length, contributing to our understanding of light curve analysis in astrophysics.


\begin{table}[ht]
\centering
\caption{Model Parameters for Different Sizes. The depth refers to the number of Transformer blocks it contains. The head denotes the number of attention heads in the multi-head self-attention mechanism within each Transformer block. The width represents the dimensionality of the hidden layers in each Transformer block.}

\begin{tabular}{@{}llll@{}}
\toprule
\label{size}
\textbf{Parameter (Million)} & \textbf{Depth} & \textbf{Head} & \textbf{Width} \\
\midrule
85 (GPT-2, default) & 12 & 12 & 768 \\
308 (medium) & 24 & 16 & 1024 \\
708 (large) & 36 & 20 & 1280 \\
1477 (XLarge) & 48 & 25 & 1600 \\
\bottomrule
\end{tabular}
\label{tab:model_params}
\end{table}

\section{Model Evaluation on Scientific Tasks} \label{scientific tasks}

\begin{figure}[!htbp]
    \centering
    \includegraphics[width=0.7\textwidth]{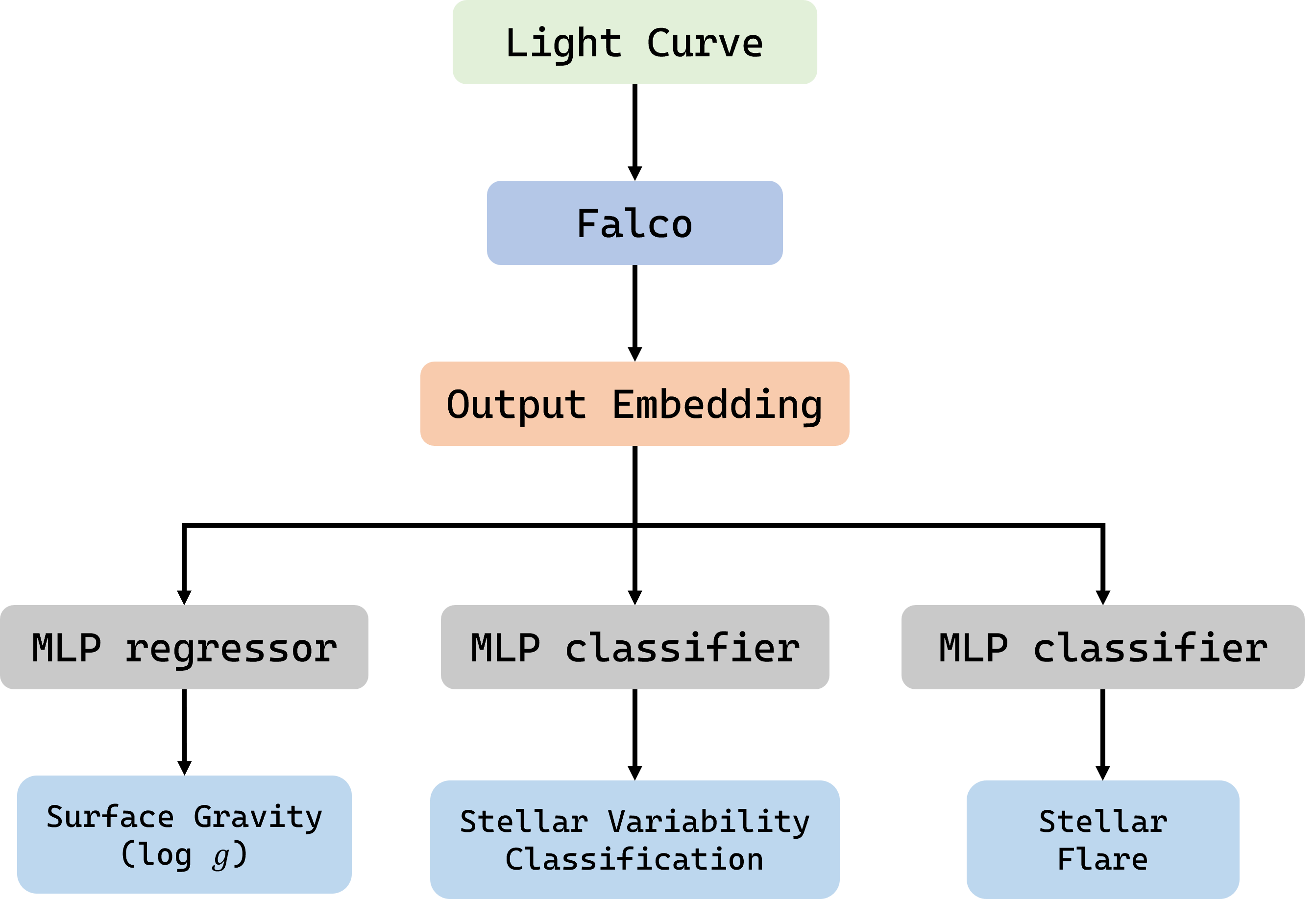} 
    \caption{Architecture diagram of foundation models used to represent different scientific task models.}
    \label{fig:down}
\end{figure}

This section presents a detailed evaluation of FALCO through three typical light curve analysis tasks in time domain astronomy: stellar variability classification, surface gravity (log $g$) estimation, and stellar flare identification. To validate the effectiveness and generalizability of FALCO on different tasks, we first use the model to encode light curve data into latent representations. As shown in Figure \ref{fig:down}, the output embeddings are subsequently fed into task-specific MLPs, with hyperparameters optimized through five-fold cross-validation, providing a quantitative assessment of their effectiveness in specific tasks. The specific configurations of the three MLP models are presented in Table \ref{table:mlp-params}. Results demonstrate outstanding performance on the three scientific tasks.The following subsections present detailed performance metrics and scientific implications for each task.

\begin{table}[!ht]
\centering

\caption{Comparison of MLP Model Parameters}
\begin{tabular}{llll}
\hline
\textbf{Parameter} & \textbf{Classification} & \textbf{log $g$ estimation} & \textbf{Flare identification} \\ \hline
Solver  & L-BFGS-B (\texttt{lbfgs}) & L-BFGS-B (\texttt{lbfgs}) & SGD (\texttt{sgd}) \\
Learning Rate Schedule & Inverse Scaling & Inverse Scaling & Adaptive (\texttt{adaptive}) \\
Initial Learning Rate  & 0.0001 & 0.001 & 0.01 \\
Hidden Layer Sizes & (80) & (100, 50, 20) & (1024, 256, 64) \\
L2 Regularization Parameter & 0.001 & 0.001 & 0.001 \\
Activation Function & Hyperbolic Tangent (\texttt{tanh}) & Hyperbolic Tangent (\texttt{tanh}) & Hyperbolic Tangent (\texttt{tanh}) \\
Early Stopping  & Enabled & Enabled & Enabled \\
 \hline
\end{tabular}
\label{table:mlp-params}
\end{table}

\subsection{Stellar Variability Classification}

\begin{figure}[htbp]
    \centering
    \includegraphics[width=0.5\textwidth]{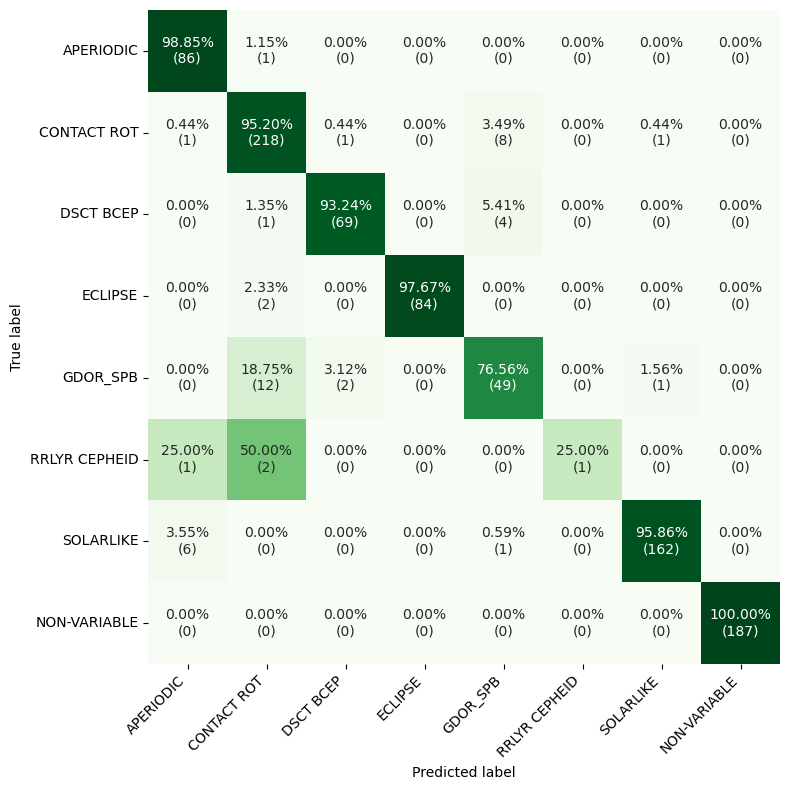} 
    \caption{Confusion matrix for stellar variability classification in the context of XLarge-200, where the labeling scheme differs by including the NON-VARIABLE class and excluding the INSTRUMENT class. Each cell similarly presents the percentage and absolute number of instances, with darker diagonal cells indicating higher classification accuracy. Compared to Figure \ref{fig:class cm2}, the introduction of the NON-VARIABLE category shifts some borderline cases and slightly alters the distribution of misclassifications among closely related classes.}
    \label{fig:class cm1}
\end{figure}

In the stellar variability classification tasks, the evaluation metrics employed were accuracy, with output reports including precision, recall, and F1 score. For this task, using an input length of 200 data points from light curves and a XLarge parameter scale model as an example, the optimal configuration of the MLP model achieved a classification accuracy of 95\%. The confusion matrix illustrating the performance of the eight-class classification task is depicted in Figure \ref{fig:class cm1}. Even in the large-scale setting, a classification accuracy of 90\% is achieved with input sequences of length 50, highlighting the effectiveness of our approach in leveraging shorter input sequences to enhance model performance. The classification performance under varying model parameter scales is summarized in Table \ref{tab:classification}. Detailed classification reports are also provided in Table \ref{tab:classification-report}, offering insights into precision, recall, F1 score, and support for each class.

\begin{figure}[htbp]
    \centering
    \includegraphics[width=0.5\textwidth]{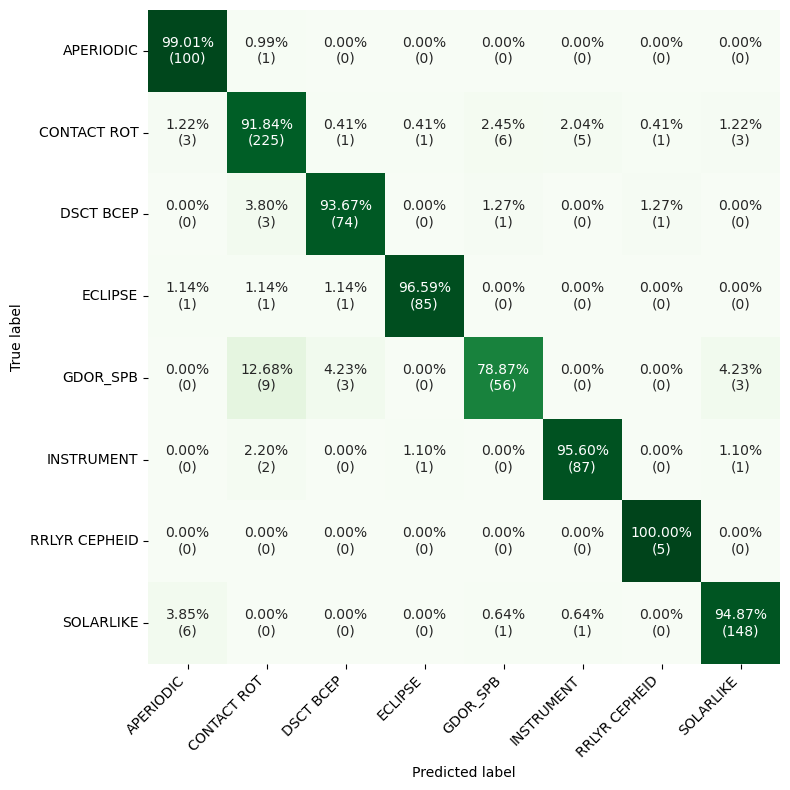} 
    \caption{Confusion matrix for stellar variability classification in the context of Large-200 under the labeling scheme proposed by \cite{audenaert2021tess}. This scheme includes the INSTRUMENT class while excluding the NON-VARIABLE class, facilitating direct comparison with the results reported in \cite{audenaert2021tess}. The true labels (rows) are compared against the predicted labels (columns), with each cell displaying both the percentage and the absolute number of instances.}
    
    \label{fig:class cm2}
\end{figure}

To facilitate comparison with the work of \cite{audenaert2021tess}, the same labeling scheme was adopted, which includes the instrument class but excludes the non-variable class. Under the labeling conditions set by \cite{audenaert2021tess}, the classification accuracy reached 95\% in the context of large-200. The confusion matrix for this scenario is presented in Figure \ref{fig:class cm2}. This classification performance surpasses that of the individual classifiers (approximately 91\%) and the metaclassifier (94.9\%) reported in \cite{audenaert2021tess}. It should be noted that their 94.9\% accuracy was obtained with a 9-class setup including the constant class, which consists of randomly generated noise, whereas our evaluation uses the same 8-class scheme without the constant class.

Although both Figures \ref{fig:class cm1} and Figures  \ref{fig:class cm2} exhibit a classification accuracy of 95\%, the performance of the “RRLYR CEPHEID” class in Figure \ref{fig:class cm1} is somewhat inferior. This class represents the smallest proportion of the dataset, accounting for only 0.73\% of the total samples. The differences in classification performance can be attributed to variations in labeling schemes and the characteristics of the datasets utilized. With only 62 samples, the limited representation of the “RRLYR CEPHEID” class likely hinders the model's ability to accurately identify this category, resulting in increased confusion with other classes, such as “CONTACT ROT” and “ECLIPSE.”

\begin{table}[htbp]
\caption{Accuracy of stellar variability classification in different contexts}
\centering 
\begin{tabular}{lcccc} 
    \toprule 
    Size & XLarge & large & medium & default\\
    
     Input Length& 200 & 200  & 200 & 200  \\
    \midrule
    Accuracy  & 0.951 & 0.950 & 0.944 & 0.942 \\
    \bottomrule
\end{tabular}

\label{tab:classification} 
\end{table}

\begin{table}[ht]
\centering
\caption{Report of stellar variability classification in the context of Xlarge-200}
\label{tab:classification-report}
\begin{tabular}{@{}lcccc@{}}
\toprule
\multicolumn{1}{c}{XLarge-200} & precision & recall & f1-score & support \\
\midrule
APERIODIC & 0.91 & 0.99 & 0.95 & 87 \\
CONTACT\_ROT & 0.92 & 0.95 & 0.94 & 229 \\
DSCT\_BCEP & 0.96 & 0.93 & 0.95 & 74 \\
ECLIPSE & 1.00 & 0.98 & 0.99 & 86 \\
GDOR\_SPB & 0.79 & 0.77 & 0.78 & 64 \\

RRLYR\_CEPHEID & 1.00 & 0.25 & 0.40 & 4 \\
SOLARLIKE & 0.99 & 0.96 & 0.97 & 169 \\
non-variable  & 1.00 & 1.00 & 1.00 & 187 \\
\midrule
accuracy & \multicolumn{3}{c}{0.95} & 900 \\
macro avg & 0.95 & 0.85 & 0.87 & 900 \\
weighted avg & 0.95 & 0.95 & 0.95 & 900 \\
\bottomrule
\end{tabular}
\end{table}

\subsection{Surface Gravity Estimation}

\begin{figure}[htbp]
    \centering
    \includegraphics[width=0.8
    \textwidth]{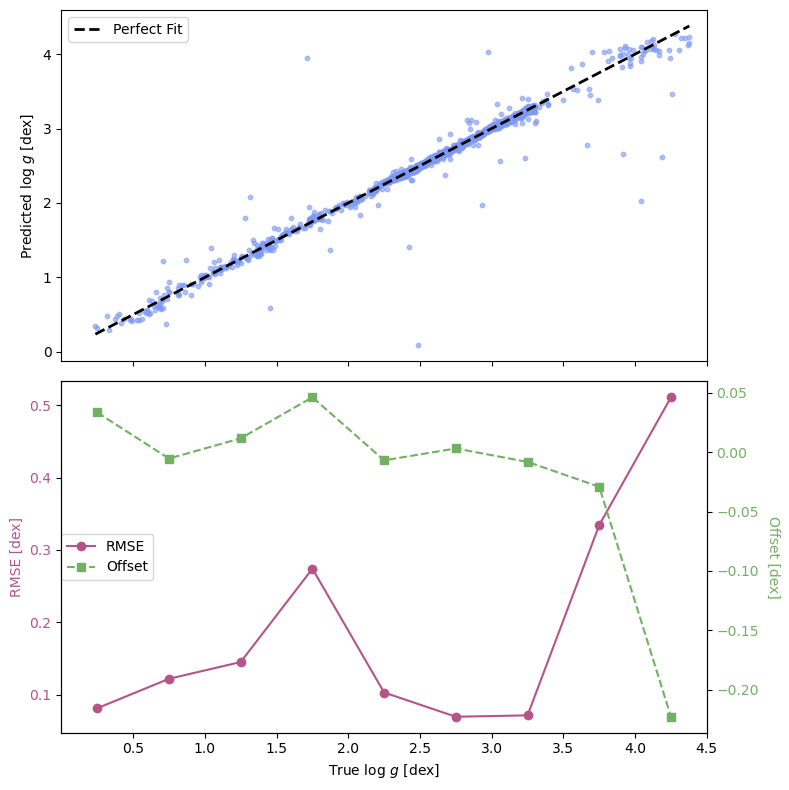} 

    \caption{Comparison of predicted and true surface gravity values (log $g$) using FALCO. The top panel shows the predicted log g values (y-axis) plotted against the true log $g$ values (x-axis), with the dashed black line representing a perfect fit. The scatter of points illustrates the accuracy of the predictions, with the majority clustering around the perfect fit line. The bottom panel provides a quantitative assessment of the prediction performance across different log g intervals. The purple line represents RMSE in dex, indicating the average prediction error for each interval, while the green dashed line shows the offset (mean prediction bias) for the same intervals. The offset values reveal systematic deviations. This figure demonstrates the effectiveness of FALCO in surface gravity estimation and highlights areas where the model performs robustly or exhibits potential biases.}
    \label{fig:logg}
\end{figure}

For the surface gravity (log $g$) estimation task, the evaluation metrics utilized included the coefficient of determination ($R^2$), mean squared error (MSE), mean absolute error (MAE), root mean squared error (RMSE), and the standard deviation ($\sigma$) of the error distribution. In the context of this study, RMSE and $\sigma$ are considered equivalent. Below is a detailed description of these metrics:

\begin{itemize}
\item \textbf{$ R^2 $}: It measures the proportion of variance in the dependent variable that can be predicted from the independent variables. It quantifies the degree to which the model's predictions correlate with the actual values.
    
\item \textbf{MSE}: It measures the average of the squares of the differences between the predicted and actual values. It is particularly sensitive to larger errors due to the squaring operation.
    
\item \textbf{MAE}: It measures the average of the absolute differences between the predicted and actual values. Unlike MSE, it treats all errors equally and does not amplify the effect of larger errors.
    
\item \textbf{RMSE and the standard deviation ($\sigma$) of the error distribution}: RMSE is the square root of MSE, restoring the error measurement to the original units of the data. It retains the sensitivity of MSE to larger errors while providing a more interpretable scale. $\sigma$ measures the dispersion of the prediction errors, indicating the deviation of the errors from their mean. When the error distribution is assumed to be normal, $\sigma$ is equivalent to RMSE.
\end{itemize}

In the task of surface gravity (log $g$) estimation using an input length of 200 data points from light curves and employing a model with a XLarge parameter scale, we achieved an RMSE/$\sigma$ of 0.1305 and an MSE of 0.363, without applying any clipping or outlier removal during performance evaluation. Specifically, our model attained an RMSE of approximately 0.02 dex at log $g = 3$, achieved RMSE $<$ 0.08 dex in the luminous red giant region (log $g < 1$), and exhibited an RMSE of about 0.2-0.25 dex for log $g > 3.5$. These results demonstrate a clear advantage over previous studies. SWAN \citep{sayeed2021swan} reported an RMSE of approximately 0.03\,dex at log $g \approx 3$, which increased to 0.3--0.5\,dex in the subgiant and upper red giant regions. In contrast, our model achieved better performance in these regions, maintaining RMSE below 0.06\,dex for log $g < 1$. Additionally, Astroconformer \citep{pan2024astroconformer} achieved an RMSE of 0.017\,dex in data-rich regions (log $g \approx 3$) and maintained RMSE $<$ 0.1\,dex in both the low-gravity (log $g < 1$) and high-gravity (log $g > 4$) regimes. It is worth noting that these results were obtained using a dataset comprising approximately 14,000 light curves. In contrast, our model achieved a comparable RMSE of 0.02\,dex at log $g \approx 3$, despite being trained on a significantly smaller dataset of only around 6,000 light curves and undergoing only a single epoch of training within the foundation model framework. Furthermore, compared to the self-supervised model presented in \citet{pan2024scaling}, which achieved an MSE of 0.059 with $10^4$ training samples and multiple training epochs, our model demonstrates superior performance with fewer training data, achieving a lower MSE/RMSE. Figure \ref{fig:logg} further illustrates the RMSE variations across different log $g$ intervals, confirming the stability and effectiveness of our model across a wide parameter range.

The results indicate that the parameter estimation of log $g$ performs best in the 2-3 dex range, while the estimation in the 0.5-2 dex and 3.5-4 dex ranges shows some deficiencies. As illustrated in Figure \ref{fig:loghist}, the majority of stars within the 2-3 dex range are primarily red clump stars, which are abundant in number. Figure \ref{fig:logg} further demonstrates that the fitting of log $g$ parameters in the 2-3 dex range is superior. A larger sample size aids in statistical analysis, enhancing the effectiveness of model training and reducing the impact of random fluctuations on the results. Stars in the 2-3 dex range are predominantly in specific stages of stellar evolution, where their physical properties are relatively stable and their variation patterns are clearer. This stability makes predictions based on physical models or empirical relationships more reliable. The 0.5-2 dex range mainly comprises luminous red giant stars, while the 3.5-4 dex range includes main sequence stars, which exhibit minimal individual differences, making it difficult to accurately describe subtle variations with existing models. Compared to the 2-3 dex range, stars in the 0.5-2 dex and 3.5-4 dex ranges may be less numerous or possess lower brightness, limiting the availability of high-quality observational data and consequently restricting the effective sample size for model training, thereby impacting predictive performance. The differences observed in fitting effectiveness and RMSE within the log $g$ ranges reflect the inherent distinctions among stellar populations and our current understanding of these differences. With the accumulation of more high-quality observational data and advancements in technology, it is anticipated that predictive accuracy across all ranges will improve in the future.

Similarly, in the surface gravity (log $g$) estimation task, we compared the performance of using MLP to process the original light curves, which also yielded poor results, as shown in Table \ref{tab:comparison}. This observation led us to conclude that simple methods like MLP are not suitable for performing parameter regression tasks directly on time-series data. 

After validating the model's performance, we analyzed the output embeddings of all Kepler light curves in the dataset to derive the log $g$ parameters for the entire dataset. This data product will also be released for reference by researchers. The surface gravity parameters (log g) derived from the analysis of the complete Kepler dataset using the FALCO model are available for download and analysis at https://nadc.china-vo.org/res/paperdata/\footnote{\url{https://nadc.china-vo.org/res/paperdata/}}.

Additionally, we conducted comparative analyses on the effects of varying model parameter scales  for this task. As shown in Table \ref{tab:performance_metrics}, it is evident that within this task framework, larger model parameter scales contribute to superior performance.

    

\begin{table}[!htbp]
    \centering
    \caption{Performance of surface gravity (log $g$) task in different contexts}
    \begin{tabular}{lccccc}
        \toprule 
        & XLarge & large & medium & default\\
        
        & 200 & 200  & 200  & 200  \\
        \midrule
        $R^2$  & 0.9561 & 0.9461 & 0.9374 & 0.9337 \\
        MSE   & 0.0170 & 0.0209 &  0.0243 & 0.0257 \\
        MAE  & 0.0363 & 0.0347 & 0.0522 & 0.0447 \\
        RMSE($\sigma$)  & 0.1305 & 0.1446 & 0.1558 & 0.1604\\
        \bottomrule
    \end{tabular}

    \label{tab:performance_metrics}
\end{table}

\subsection{Stellar Flare Identification}

\begin{figure}[htbp]
    \centering
    \includegraphics[width=0.4\textwidth]{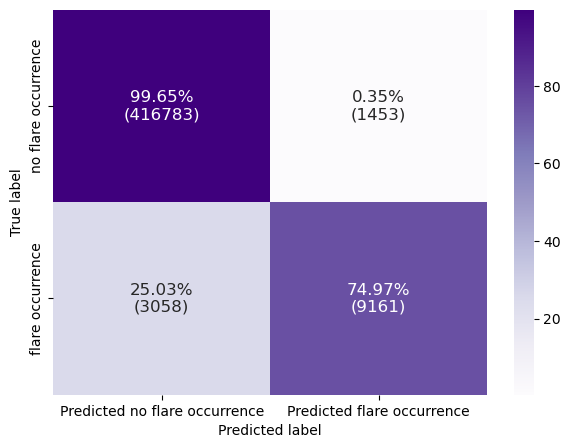} 
    \caption{Confusion matrix illustrating the performance of the flare identification model. The matrix compares the predicted labels (x-axis) with the true labels (y-axis) for two classes: "no flare occurrence" and "flare occurrence." }
    \label{fig:flarecm}
\end{figure}

In the task of flare identification, we frame the problem as a binary classification task to determine whether a flare event occurs. Our primary focus, however, is on the outcomes related to the identification of flare events (class 1). As with the classification task, the evaluation metrics include accuracy, precision, recall, and F1 score. With an input length of 200, the model achieves a precision of 87\% and a recall of 76\% for class 1 (flare occurrence). For segments of light curves classified as class 0 (no flare occurrence), the model demonstrates exceptional performance, achieving a precision of 99\% and a recall of 100\%. Figure \ref{fig:flarecm} shows the confusion matrix for flare identification, while Table \ref{tab:flare} presents the performance metrics.

Given that flares are rare events, minimizing false positives is crucial; therefore, we place greater emphasis on the F1 score and precision of flare predictions. However, we believe that the requirements for the false positive rate vary across different flare identification tasks. Some identifications aim to minimize the false positive rate, while others prioritize reducing the false negative rate. In such cases, the model's decision threshold should be adjusted according to the specific circumstances.

\begin{table}[h]
\centering
\caption{Performance of stellar flare identification task}
\label{tab:flare}

\begin{tabular}{ccccc}
\toprule
\textbf{Condition} & \textbf{Label} & \textbf{Precision} & \textbf{Recall} & \textbf{F1-score} \\
\midrule
\textbf{output embedding by FALCO} & No flare occurrence & 0.99 & 1.00 & 0.99 \\
& Flare occurrence & 0.87 & 0.76 & 0.80 \\
\midrule
\textbf{original light curves without FALCO} & No flare occurrence & 0.95 & 0.98 & 0.96 \\
& Flare occurrence & 0.83 & 0.72 & 0.77 \\
\bottomrule
\end{tabular}

\end{table}

This study currently focuses on the binary classification task for flare identification. We plan to extend the method to the prediction of flares in future work.

\section{Discussion}\label{disscussion}


In this section, we present a systematic analysis and discussion of the key factors related to the performance of FALCO, providing insights on model efficacy, design considerations, and potential optimizations. Specifically, our analysis focuses on three key aspects: the effectiveness of learned representations, the effects of varying input sequence length, and the performance implications of model scaling. These findings not only illuminate the underlying mechanisms of FALCO's effectiveness but also inform directions for future refinements. 

\subsection{Efficacy of Foundation Models and Light Curve Representations}
 
In this study, the efficacy of foundation models in addressing the challenges of light curve data analysis is demonstrated. The findings confirm the potential of Transformer architectures and self-supervised learning in processing light curve data. Specifically, it is shown that these advanced methodologies can effectively capture the intricate patterns within light curve datasets, thereby enhancing the accuracy and efficiency of astronomical observations and analyses.

This work underscores the advantages of self-supervised learning, which mitigates the reliance on large annotated datasets by leveraging the inherent structure of the data itself for pre-training. This approach not only reduces the burden of data labeling but also broadens the applicability of machine learning models to scenarios where labeled data are scarce or difficult to obtain. Overall, our research highlights the improved impact of foundation models on the field of astrophysics.

To further analyze and validate the efficacy of foundation models and learned representations, we conducted ablation studies for each scientific task. These experiments aimed to compare the outcomes of directly using original light curves without FALCO against those obtained through FALCO, which provided output embeddings. The results, as shown in the Table \ref{tab:flare} and Table \ref{tab:comparison}, indicate that utilizing light curves directly for scientific tasks via MLP presents certain limitations. In contrast, the output embeddings derived from FALCO demonstrate noticeably improved performance. This finding further corroborates the effectiveness of learned representations of self-supervised foundation model in enhancing the analysis of light curve data.

Additionally, we compared the classification performance of the original light curves without FALCO processing using MLPs, which yielded an accuracy of only 0.245. We also employed other machine learning methods, such as Random Forest, XGBoost, and PCA, all of which produced poor results. The surface gravity (log $g$) estimation task yielded similarly unsatisfactory outcomes. In the stellar flare identification task, the improvement in performance when using output embeddings from FALCO compared to the original light curves without FALCO was not substantial, with flare identification results increasing from 83\% to 87\%. This limited enhancement can be attributed to the nature of the task, as the test data consisted of pre-segmented 200-length clips where flares are already quite pronounced. In contrast, the datasets for the stellar variability classification and surface gravity (log $g$) estimation tasks involve long-term observations with more intricate underlying patterns that are not immediately discernible. 

\begin{table}[htbp]
    \centering
    \caption{Comparison of the performance of Original Light Curves Without FALCO and Output Embedding by FALCO under 200 input length and XLarge scale model for the tasks of Stellar Variability Classification and Surface Gravity (log $g$) Estimation.}
    \begin{tabular}{lcccccc}
        \toprule
        \textbf{Condition} & \multicolumn{3}{c}{\textbf{Stellar Variability Classification}} & \multicolumn{3}{c}{\textbf{Surface Gravity (log $g$) Estimation}} \\
        \cmidrule(lr){2-4} \cmidrule{5-7}
                  & \textbf{Precision}  & \textbf{Recall}   & \textbf{F1-Score}    & \textbf{$R^2$}    & \textbf{MAE}     & \textbf{RMSE}   \\
        \midrule
        \textbf{Original Light Curves Without FALCO} & 0.17  & 0.18  & 0.18  & -0.2435  & 0.4934  & 0.6947 \\
        \textbf{Output Embedding by FALCO}           & 0.95  & 0.93  & 0.94  & 0.9561   & 0.0363  & 0.1305 \\
        \bottomrule
    \end{tabular}
    
    \label{tab:comparison}
\end{table}

\begin{figure}[htbp]
    \centering
    \includegraphics[width=0.8\textwidth]{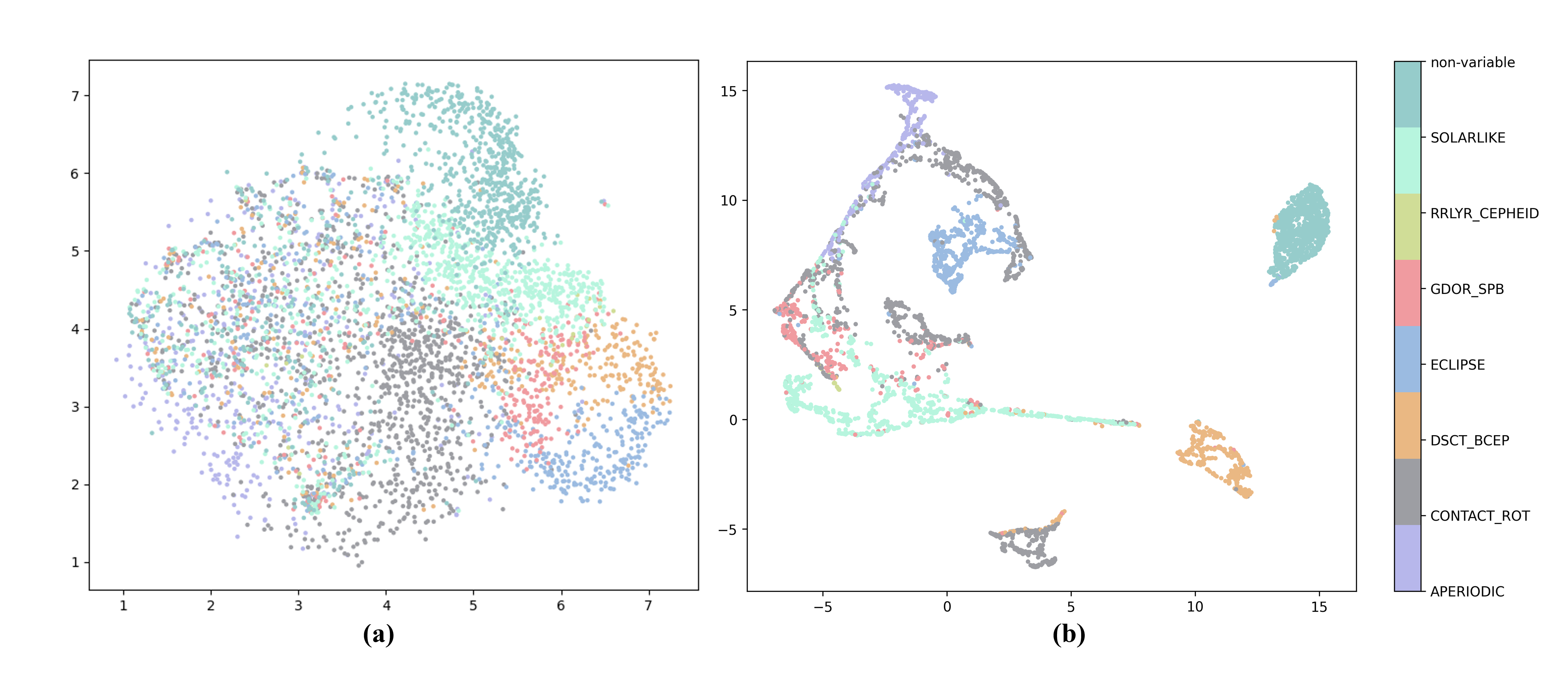} 
    \caption{UMAP Visualization Comparison Original Light Curve and Output embedding for task of stellar variability classification. Figure (a) on the left illustrates the original light curve data, while Figure (b) on the right illustrates the output embedding generated by the FALCO for classification purposes.}
    \label{fig:umap_class}
\end{figure}

\begin{figure}[htbp]
    \centering
    \includegraphics[width=0.8\textwidth]{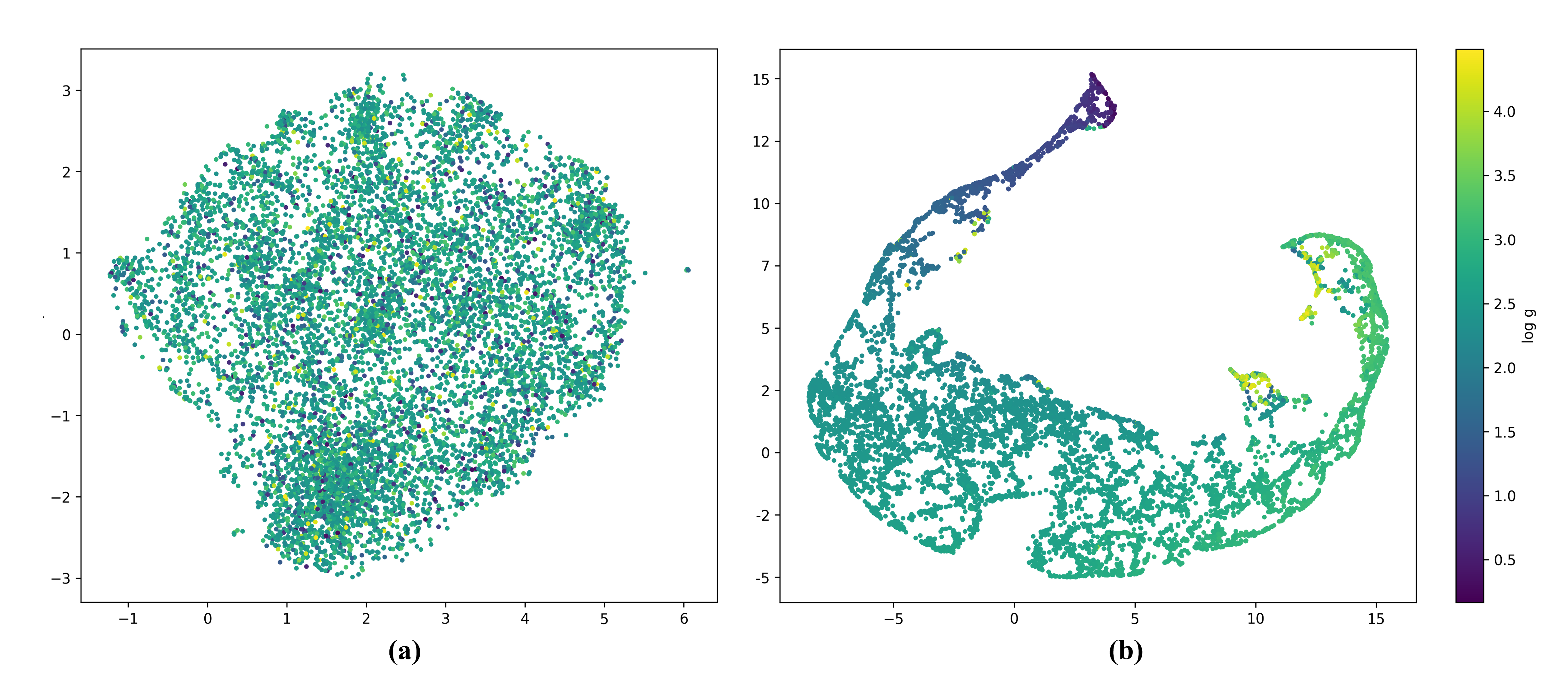} 
    \caption{UMAP Visualization Comparison Original Light Curve and Output embedding for task of surface gravity estimation. Figure (a) on the left illustrates the original light curve data, while Figure (b) on the right illustrates the output embedding generated by the FALCO for log $g$ estimation purposes.}
    \label{fig:umap_loggs}
\end{figure}

We also conducted an analysis comparing the original light curves and the output embeddings after applying Uniform Manifold Approximation and Projection (UMAP) dimensionality reduction \citep{mcinnes2018umap}. Figure \ref{fig:umap_class} illustrates the UMAP visualization comparison of the original light curves before and after decoding by FALCO for task of stellar variability classification. In panel (a) of Figure \ref{fig:umap_class}, which depicts the original light curves, distinct clusters are only vaguely discernible, with significant overlap between classes. This overlap suggests that directly applying an MLPClassifier to the raw data would result in poor classification performance, achieving an accuracy of approximately 24.5\%. In contrast, panel (b) of Figure \ref{fig:umap_class} shows a notable improvement in class separation for the output embeddings generated by FALCO. The clusters are well-defined with clear boundaries, leading to a noticeably enhanced classification accuracy of 95\%. This visual enhancement highlights the effectiveness of FALCO in improving the discriminative power of the embedded features, thereby noticeably boosting the model's classification performance. Figure \ref{fig:umap_loggs} provides a comparative UMAP visualization of the original light curves before and after processing by FALCO for task of surface gravity estimation. In panel (a) of Figure \ref{fig:umap_loggs}, representing the original light curves, the clusters are not clearly distinguishable due to overlap between different classes. This overlap indicates that estimating the log $g$ value directly from raw data would be challenging, resulting in a high RMSE of 0.6947. Conversely, panel (b) of Figure \ref{fig:umap_loggs} demonstrates a significant improvement in cluster separation within the output embeddings post-processing. The clusters exhibit clear boundaries, leading to a markedly lower RMSE of 0.1305. This improved visual representation underscores the efficacy of FALCO in enhancing feature discriminability, thus greatly improving the accuracy of log $g$ value estimation.

\subsection{Impact of Input Sequence Length}\label{length}

\begin{figure}[htbp]
    \centering
    \includegraphics[width=0.7\textwidth]{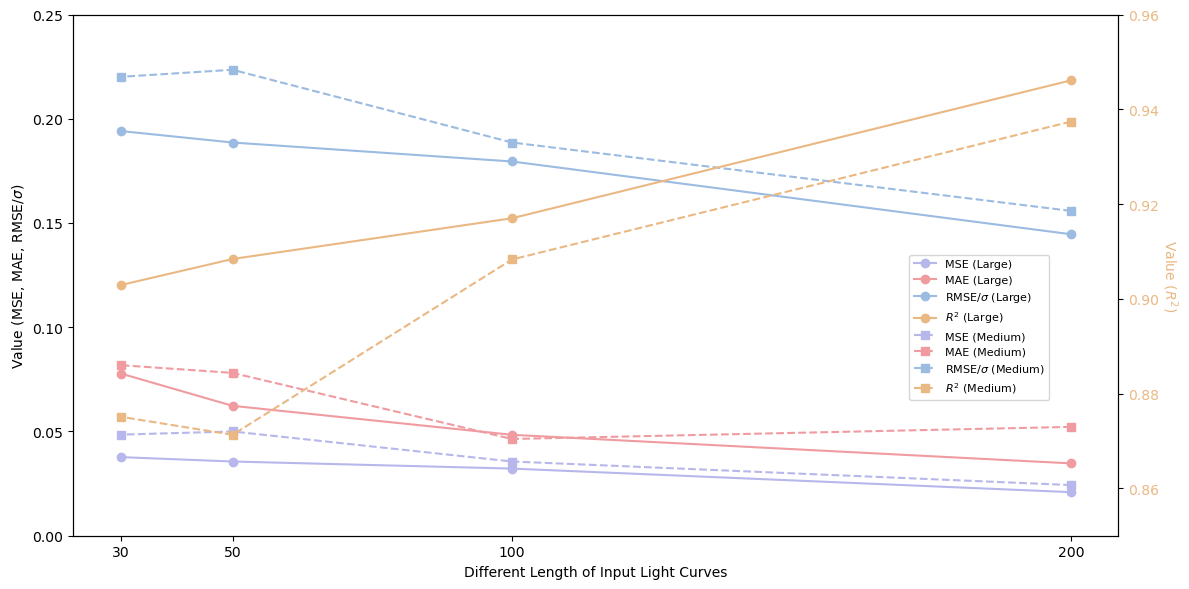} 
    \caption{This figure presents the performance of four evaluation metrics for the log $g$ estimation task, involving two different model parameter scales and four distinct light curve input lengths. The horizontal axis indicates the four input lengths, with solid lines representing the large-scale model and dashed lines indicating the medium-scale model. Four different colors are used to denote the various evaluation metrics.}
    \label{fig:length}
\end{figure}

\begin{figure}[htbp]
    \centering
    \includegraphics[width=0.7\textwidth]{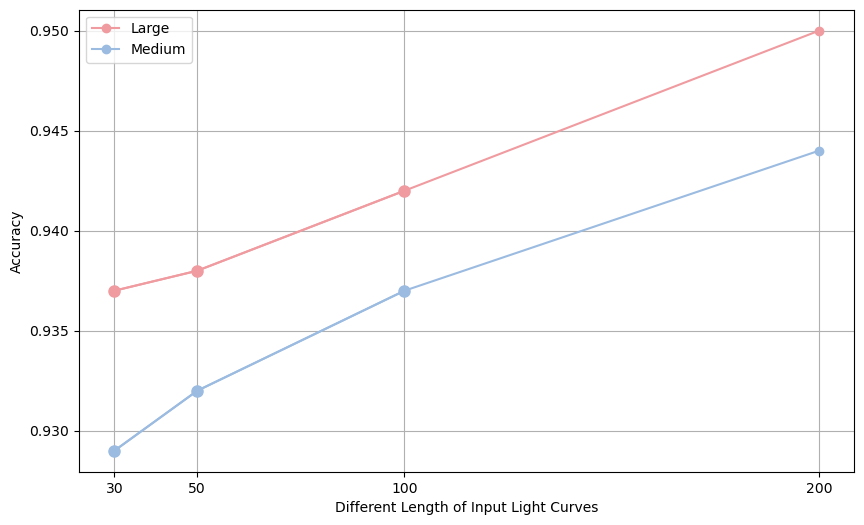} 
    \caption{Classification accuracy of models with medium and large parameter scales as a function of the input light curve length for task of stellar variability classification. The x-axis represents the length of the input light curves, while the y-axis shows the corresponding classification accuracy. The red line indicates the performance of the model with a large parameter scale, and the blue line represents the performance of the model with a medium parameter scale. The plot demonstrates that accuracy improves consistently with increasing input light curve length for both models. However, the model with a large parameter scale achieves higher accuracy across all input lengths compared to the medium-scale model.}
    \label{fig:class_result}
\end{figure}

In this study, we delve into the impact of varying lengths of light curve inputs on model performance. To comprehensively evaluate this effect, we conducted a comparative analysis using input lengths of 30, 50, 100, and 200 time steps. This range was selected to span from minimal to substantial temporal coverage, enabling quantitative assessment of how sequence length influences model accuracy across various scientific tasks based on light curves.
 
Results demonstrate a significant positive correlation between input sequence length and prediction accuracy across all measured parameters. As shown in Figure \ref{fig:length}, using surface gravity (log $g$) estimation as an example case, performance metrics exhibit a monotonic improvement with increasing sequence length. The model trained on 200 time steps achieved the highest precision. The model trained on 200 time steps achieved the highest precision, with RMSE decreasing by 0.0495 and 0.0643 compared to the 30-time-step model for large and medium model sizes, respectively. This trend is consistent across all stellar classification tasks, as evidenced in Figure \ref{fig:class_result}. Classification accuracy improved from 93.7\% to 95.0\% for the large model size and from 92.9\% to 94.4\% for the medium model size when increasing from 30 to 200 time steps.

Importantly, this relationship between sequence length and performance persisted regardless of model parameter scale. Both large-scale and medium-scale model configurations demonstrated comparable relative improvements with increased input length, suggesting that the temporal information content, rather than model complexity, is the primary determinant of performance in these tasks.

Despite the clear advantages of longer sequences, shorter light curves retain practical utility in specific operational contexts. When rapid analysis is required—such as during initial target selection or real-time telescope scheduling decisions—30-50 time step models provide sufficient discriminatory power while minimizing computational overhead. These shorter sequences enable preliminary classification with acceptable accuracy when immediate decisions are required before complete observations are available.

Thus, while short sequences can provide initial assessments, the accumulation of more data enables models based on longer sequences to deliver more reliable and accurate outcomes. Therefore, the choice of appropriate input length should be guided by the specific requirements of the application context: prioritizing longer input sequences when high precision and detailed understanding are desired, and opting for shorter sequences when rapid response and preliminary assessment are needed. This discovery provides theoretical support for future astrophysical data analysis and offers valuable insights for time-series modeling in other domains.

\subsection{Analysis of Model Scaling} \label{sclaing law}

In order to evaluate the impact of the model parameter scale on the learning ability and predictive performance in light curve analysis task, we conducted a comparative analysis using  FALCO models with parameters ranging from 85 million to 1,477 million parameters, as detailed in Tables \ref{tab:classification} and Tables \ref{tab:performance_metrics}. The results of both classification tasks and surface gravity (log $g$) estimation consistently demonstrate that larger model parameter scales yield superior performance on scientific tasks. 

As shown in Figure \ref{fig:val loss}, the validation loss during training decreases as the model parameter size increases for models of different scales. For the stellar variability classification task, accuracy decreased from 94.2\% to 95.1\%, and for the surface gravity (log $g$) estimation task, RMSE decreased This indicates that models with greater parameter scales are capable of capturing more intricate features and patterns within light curves, such as variations in brightness and periodic signals, which may contribute to improved predictive accuracy and generalization capabilities. Figure \ref{fig:class_result} also illustrates the classification accuracy under medium and large scales model. The large model with 708 million parameters shows an overall improvement in classification accuracy compared to the medium model with 308 million parameters, regardless of the light curve input lengths. The performance comparison between large-scale and medium-scale models in the surface gravity (log $g$) estimation task, shown in Figure \ref{fig:length}, further supports this finding.

This observation aligns with the Scaling Law in artificial intelligence \citep{hestness2017deep,kaplan2020scaling}, which posits that increases in model size, data quantity, and computational power tend to proportionally improve model performance. The enhanced performance of larger models can be attributed to their increased capacity to represent more sophisticated functions and relationships within the data. With a greater number of parameters, these models can better approximate the non-linearities present in light curves, leading to finer-grained feature extraction and more accurate predictions. As shown in Table \ref{tab:model_time}, the training time per epoch varies among models of different sizes, with the XLarge model taking the longest time at 20.3 hours. However, it is important to note that while larger models offer performance benefits, they also come with higher computational costs and potential challenges related to overfitting, especially if the dataset size does not scale accordingly. 

In summary, our experiments underscore the importance of model scale in achieving optimal performance for time series analysis tasks, such as those encountered in astrophysics. They provide strong empirical support for the Scaling Law and suggest that future research should continue to explore the balance between model size, data availability, and computational resources to maximize predictive power while maintaining efficiency.

\begin{table}[h]
\centering
\caption{Training Time per Epoch for Different Model Sizes}
\label{tab:model_time}
\begin{tabular}{lcccc}
\toprule
& \textbf{Default} & \textbf{Medium} & \textbf{Large} & \textbf{XLarge} \\
\midrule
\textbf{Parameter (Million)} & 85 & 308 & 708 & 1477 \\
\textbf{Training Time per Epoch (hours)} & 1.9 & 4.8 & 10.6 & 20.3 \\
\bottomrule
\end{tabular}
\end{table}


\section{Application} \label{application}

\begin{figure}[htbp]
    \centering
    \includegraphics[width=0.7\textwidth]{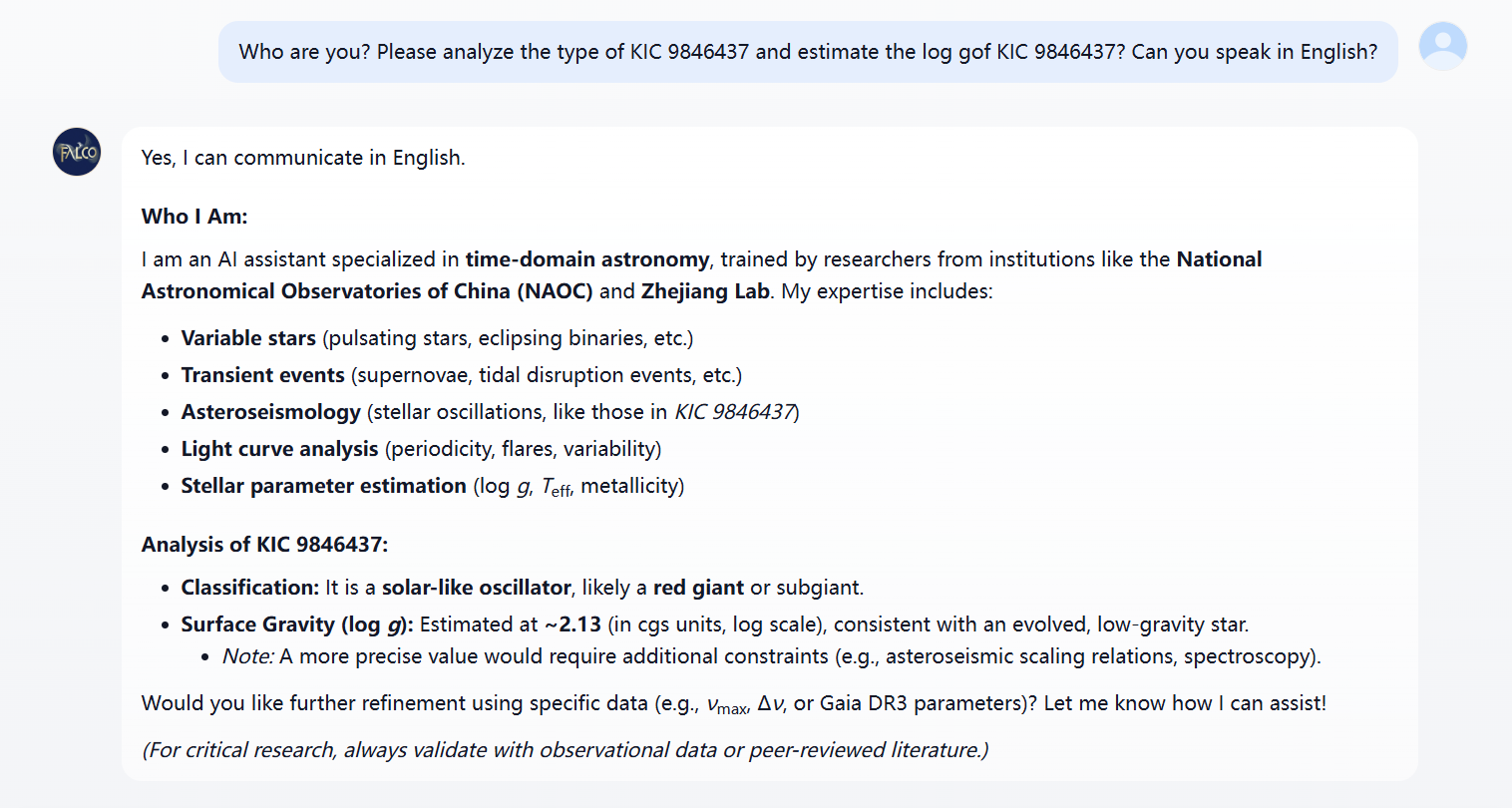} 
    \caption{Q\&A example of FALCO deployment in the NADC portal}
    \label{fig:talk}
\end{figure}

\begin{figure}[htbp]
    \centering
    \includegraphics[width=0.9\textwidth]{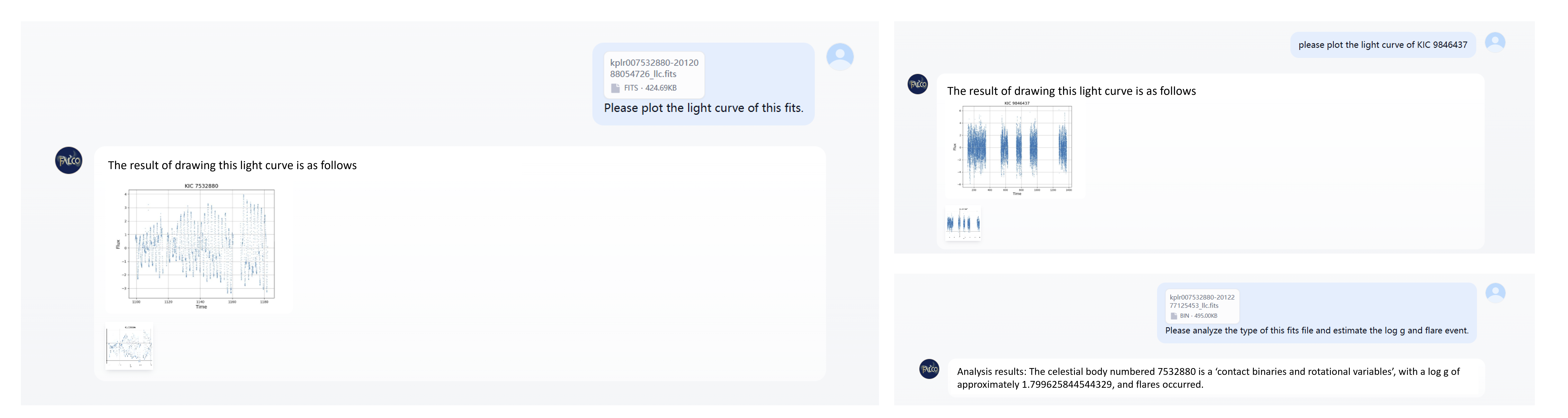} 
    \caption{Examples of light curves and scientific task analysis, including light curves generated from uploaded FITS files or KIC numbers, and a detailed analysis of scientific tasks based on these data.}
    \label{fig:talk2}
\end{figure}

The FALCO model has been deployed with an LLM web interface to enable interactive analysis of light curves. This service is accessible through the NAOC astronomical model platform maintained by the National Astronomical Data Center of China (NADC). The interactive interface of the FALCO application is shown in Figure \ref{fig:talk}.The system allows users to upload Kepler light curve FITS files or query specific Kepler Input Catalog (KIC) IDs to perform stellar classification, surface gravity estimation (log $g$), and flare identification, as shown in Figure \ref{fig:talk2}. Our initiative seeks to foster a collaborative environment where astronomers can leverage advanced machine learning tools to enhance their research.
By offering pre-trained models through an intuitive interface, we aim to reduce the technical barriers to applying AI techniques and models in astronomical data analysis. Future development plans include expanding the capability of the FALCO model to address additional scientific tasks, implementing automated adaptation mechanisms for user-specific targets, and open-sourcing the model to facilitate further research based on FALCO.

\section{Conclusion}\label{conclusion}

In this work, we presented FALCO, a Foundation model of Astronomical Light Curves for time-domain astronomy that leverages self-supervised learning and Transformer architectures. Our analysis demonstrates that FALCO generates embeddings that effectively capture intrinsic patterns in light curves, enabling clear discrimination between diverse variable phenomena as visualized through UMAP projections. This enhanced feature representation translates to exceptional performance across multiple scientific tasks: 95\% accuracy in 8-class classification, RMSE of 0.13 in log $g$ estimation, and 87\% precision in flare identification.

The scaling properties of FALCO align with established AI scaling laws—larger models outperform smaller ones, while longer input sequences capture richer temporal features. These characteristics make the model versatile for both high-precision analysis and preliminary assessments with limited data. FALCO has been deployed as an interactive platform that integrates large language models, providing user-friendly access to advanced analytical capabilities.


Despite these achievements, FALCO’s current implementation faces limitations in its point-based tokenization approach and fixed cadence requirements, which constrain its application to irregularly sampled data.  Future refinements will focus on developing a more flexible architecture capable of handling variable sampling rates and complex observational patterns inherent in diverse astronomical datasets. Our long-term vision extends beyond technical improvements to building a more universal foundation model that incorporates multi-facility data and addresses additional scientific applications. Through these advances, FALCO aims to serve as a cornerstone tool for the astronomical community in the era of time-domain big data, supporting and accelerating scientific discovery as observational datasets continue to grow in volume and complexity.


\begin{acknowledgments}
This work is supported by the National Key R\&D Program of China (2022YFF0711500, 2019YFA0405503), National Natural Science Foundation of China (NSFC)(12403102, 12373110, 12273077, 12103070, 12422303), Strategic Priority Research Program of the Chinese Academy of Sciences (XDB0550101, XDB0550103), Key R\&D Program of Zhejiang (2024SSYS0006). Data resources are supported by China National Astronomical Data Center (NADC), CAS Astronomical Data Center and Chinese Virtual Observatory (China-VO). This work is supported by Astronomical Big Data Joint Research Center, co-founded by National Astronomical Observatories, Chinese Academy of Sciences and Alibaba Cloud. We sincerely appreciate the computational power and resources provided by Zhejiang Lab.
\end{acknowledgments}

%

\vspace{5mm}
\facilities{Kepler}


\software{Astropy \citep{2013A&A...558A..33A,2018AJ....156..123A},
Scikit-learn
\citep{pedregosa2011scikit},
Tensorflow
\citep{martin2015tensorflow},
Pandas
\citep{mckinney2011pandas},
Numpy
\citep{harris2020array},
Matplotlib
\citep{hunter2007matplotlib},
Jupyter Notebook 
\citep{kluyver2016jupyter}
}





\bibliography{sample631}{}

\begin{thebibliography}{}
\expandafter\ifx\csname natexlab\endcsname\relax\def\natexlab#1{#1}\fi
\providecommand{\url}[1]{\href{#1}{#1}}
\providecommand{\dodoi}[1]{doi:~\href{http://doi.org/#1}{\nolinkurl{#1}}}
\providecommand{\doeprint}[1]{\href{http://ascl.net/#1}{\nolinkurl{http://ascl.net/#1}}}
\providecommand{\doarXiv}[1]{\href{https://arxiv.org/abs/#1}{\nolinkurl{https://arxiv.org/abs/#1}}}

\bibitem[{Aerts {et~al.}(2010)Aerts, Christensen-Dalsgaard, \& Kurtz}]{aerts2010asteroseismology}
Aerts, C., Christensen-Dalsgaard, J., \& Kurtz, D.~W. 2010, Asteroseismology (Springer Science \& Business Media)

\bibitem[{{Astropy Collaboration} {et~al.}(2013){Astropy Collaboration}, {Robitaille}, {Tollerud}, {Greenfield}, {Droettboom}, {Bray}, {Aldcroft}, {Davis}, {Ginsburg}, {Price-Whelan}, {Kerzendorf}, {Conley}, {Crighton}, {Barbary}, {Muna}, {Ferguson}, {Grollier}, {Parikh}, {Nair}, {Unther}, {Deil}, {Woillez}, {Conseil}, {Kramer}, {Turner}, {Singer}, {Fox}, {Weaver}, {Zabalza}, {Edwards}, {Azalee Bostroem}, {Burke}, {Casey}, {Crawford}, {Dencheva}, {Ely}, {Jenness}, {Labrie}, {Lim}, {Pierfederici}, {Pontzen}, {Ptak}, {Refsdal}, {Servillat}, \& {Streicher}}]{2013A&A...558A..33A}
{Astropy Collaboration}, {Robitaille}, T.~P., {Tollerud}, E.~J., {et~al.} 2013, \aap, 558, A33, \dodoi{10.1051/0004-6361/201322068}

\bibitem[{{Astropy Collaboration} {et~al.}(2018){Astropy Collaboration}, {Price-Whelan}, {Sip{\H{o}}cz}, {G{\"u}nther}, {Lim}, {Crawford}, {Conseil}, {Shupe}, {Craig}, {Dencheva}, {Ginsburg}, {VanderPlas}, {Bradley}, {P{\'e}rez-Su{\'a}rez}, {de Val-Borro}, {Aldcroft}, {Cruz}, {Robitaille}, {Tollerud}, {Ardelean}, {Babej}, {Bach}, {Bachetti}, {Bakanov}, {Bamford}, {Barentsen}, {Barmby}, {Baumbach}, {Berry}, {Biscani}, {Boquien}, {Bostroem}, {Bouma}, {Brammer}, {Bray}, {Breytenbach}, {Buddelmeijer}, {Burke}, {Calderone}, {Cano Rodr{\'\i}guez}, {Cara}, {Cardoso}, {Cheedella}, {Copin}, {Corrales}, {Crichton}, {D'Avella}, {Deil}, {Depagne}, {Dietrich}, {Donath}, {Droettboom}, {Earl}, {Erben}, {Fabbro}, {Ferreira}, {Finethy}, {Fox}, {Garrison}, {Gibbons}, {Goldstein}, {Gommers}, {Greco}, {Greenfield}, {Groener}, {Grollier}, {Hagen}, {Hirst}, {Homeier}, {Horton}, {Hosseinzadeh}, {Hu}, {Hunkeler}, {Ivezi{\'c}}, {Jain}, {Jenness}, {Kanarek}, {Kendrew}, {Kern}, {Kerzendorf}, {Khvalko}, {King}, {Kirkby}, {Kulkarni},
  {Kumar}, {Lee}, {Lenz}, {Littlefair}, {Ma}, {Macleod}, {Mastropietro}, {McCully}, {Montagnac}, {Morris}, {Mueller}, {Mumford}, {Muna}, {Murphy}, {Nelson}, {Nguyen}, {Ninan}, {N{\"o}the}, {Ogaz}, {Oh}, {Parejko}, {Parley}, {Pascual}, {Patil}, {Patil}, {Plunkett}, {Prochaska}, {Rastogi}, {Reddy Janga}, {Sabater}, {Sakurikar}, {Seifert}, {Sherbert}, {Sherwood-Taylor}, {Shih}, {Sick}, {Silbiger}, {Singanamalla}, {Singer}, {Sladen}, {Sooley}, {Sornarajah}, {Streicher}, {Teuben}, {Thomas}, {Tremblay}, {Turner}, {Terr{\'o}n}, {van Kerkwijk}, {de la Vega}, {Watkins}, {Weaver}, {Whitmore}, {Woillez}, {Zabalza}, \& {Astropy Contributors}}]{2018AJ....156..123A}
{Astropy Collaboration}, {Price-Whelan}, A.~M., {Sip{\H{o}}cz}, B.~M., {et~al.} 2018, \aj, 156, 123, \dodoi{10.3847/1538-3881/aabc4f}

\bibitem[{Audenaert {et~al.}(2021)Audenaert, Kuszlewicz, Handberg, Tkachenko, Armstrong, Hon, Kgoadi, Lund, Bell, Bugnet, {et~al.}}]{audenaert2021tess}
Audenaert, J., Kuszlewicz, J.~S., Handberg, R., {et~al.} 2021, The Astronomical Journal, 162, 209

\bibitem[{Baevski {et~al.}(2020)Baevski, Zhou, Mohamed, \& Auli}]{baevski2020wav2vec}
Baevski, A., Zhou, Y., Mohamed, A., \& Auli, M. 2020, Advances in neural information processing systems, 33, 12449

\bibitem[{Bommasani {et~al.}(2021)Bommasani, Hudson, Adeli, Altman, Arora, von Arx, Bernstein, Bohg, Bosselut, Brunskill, {et~al.}}]{bommasani2021opportunities}
Bommasani, R., Hudson, D.~A., Adeli, E., {et~al.} 2021, arXiv preprint arXiv:2108.07258

\bibitem[{Borucki {et~al.}(2010)Borucki, Koch, Basri, Batalha, Brown, Caldwell, Caldwell, Christensen-Dalsgaard, Cochran, DeVore, {et~al.}}]{borucki2010kepler}
Borucki, W.~J., Koch, D., Basri, G., {et~al.} 2010, Science, 327, 977

\bibitem[{Chen {et~al.}(2023)Chen, Tian, Fang, Zuo, Bird, Liu, Zhu, Zhang, Liu, \& Cui}]{chen2023discovery}
Chen, Y.-T., Tian, H.-J., Fang, M., {et~al.} 2023, Science China Physics, Mechanics \& Astronomy, 66, 299514

\bibitem[{Coughlin {et~al.}(2023)Coughlin, Bloom, Nir, Antier, Du~Laz, Van Der~Walt, Crellin-Quick, Culino, Duev, Goldstein, {et~al.}}]{coughlin2023data}
Coughlin, M.~W., Bloom, J.~S., Nir, G., {et~al.} 2023, The Astrophysical Journal Supplement Series, 267, 31

\bibitem[{Cui {et~al.}(2024)Cui, Armstrong, \& Feng}]{cui2024identifying}
Cui, K., Armstrong, D., \& Feng, F. 2024, The Astrophysical Journal Supplement Series, 274, 29

\bibitem[{Cui {et~al.}(2021)Cui, Liu, Feng, \& Liu}]{cui2021identify}
Cui, K., Liu, J., Feng, F., \& Liu, J. 2021, The Astronomical Journal, 163, 23

\bibitem[{Devlin {et~al.}(2018)Devlin, Chang, Lee, \& Toutanova}]{devlin2018bert}
Devlin, J., Chang, M.-W., Lee, K., \& Toutanova, K. 2018, arXiv preprint arXiv:1810.04805

\bibitem[{Donoso-Oliva {et~al.}(2023)Donoso-Oliva, Becker, Protopapas, Cabrera-Vives, Vishnu, \& Vardhan}]{donoso2023astromer}
Donoso-Oliva, C., Becker, I., Protopapas, P., {et~al.} 2023, Astronomy \& Astrophysics, 670, A54

\bibitem[{Dosovitskiy {et~al.}(2020)Dosovitskiy, Beyer, Kolesnikov, Weissenborn, Zhai, Unterthiner, Dehghani, Minderer, Heigold, Gelly, {et~al.}}]{dosovitskiy2020image}
Dosovitskiy, A., Beyer, L., Kolesnikov, A., {et~al.} 2020, arXiv preprint arXiv:2010.11929

\bibitem[{Gao {et~al.}(2016)Gao, Xin, Liu, Zhang, \& Gao}]{gao2016white}
Gao, Q., Xin, Y., Liu, J.-F., Zhang, X.-B., \& Gao, S. 2016, The Astrophysical Journal Supplement Series, 224, 37

\bibitem[{Graham {et~al.}(2012)Graham, Djorgovski, Mahabal, Donalek, Drake, \& Longo}]{graham2012data}
Graham, M.~J., Djorgovski, S.~G., Mahabal, A., {et~al.} 2012, Distributed and Parallel Databases, 30, 371

\bibitem[{Harris {et~al.}(2020)Harris, Millman, Van Der~Walt, Gommers, Virtanen, Cournapeau, Wieser, Taylor, Berg, Smith, {et~al.}}]{harris2020array}
Harris, C.~R., Millman, K.~J., Van Der~Walt, S.~J., {et~al.} 2020, Nature, 585, 357

\bibitem[{Hawley {et~al.}(2014)Hawley, Davenport, Kowalski, Wisniewski, Hebb, Deitrick, \& Hilton}]{hawley2014kepler}
Hawley, S.~L., Davenport, J.~R., Kowalski, A.~F., {et~al.} 2014, The Astrophysical Journal, 797, 121

\bibitem[{Hestness {et~al.}(2017)Hestness, Narang, Ardalani, Diamos, Jun, Kianinejad, Patwary, Yang, \& Zhou}]{hestness2017deep}
Hestness, J., Narang, S., Ardalani, N., {et~al.} 2017, arXiv preprint arXiv:1712.00409

\bibitem[{Hlo{\v{z}}ek(2019)}]{hlovzek2019data}
Hlo{\v{z}}ek, R. 2019, Publications of the Astronomical Society of the Pacific, 131, 118001

\bibitem[{Huber(1992)}]{huber1992robust}
Huber, P.~J. 1992, in Breakthroughs in statistics: Methodology and distribution (Springer), 492--518

\bibitem[{Hunter(2007)}]{hunter2007matplotlib}
Hunter, J.~D. 2007, Computing in science \& engineering, 9, 90

\bibitem[{Ilin {et~al.}(2019)Ilin, Schmidt, Davenport, \& Strassmeier}]{ilin2019flares}
Ilin, E., Schmidt, S.~J., Davenport, J.~R., \& Strassmeier, K.~G. 2019, Astronomy \& Astrophysics, 622, A133

\bibitem[{Ivezi{\'c} {et~al.}(2019)Ivezi{\'c}, Kahn, Tyson, Abel, Acosta, Allsman, Alonso, AlSayyad, Anderson, Andrew, {et~al.}}]{ivezic2019lsst}
Ivezi{\'c}, {\v{Z}}., Kahn, S.~M., Tyson, J.~A., {et~al.} 2019, The Astrophysical Journal, 873, 111

\bibitem[{Jin {et~al.}(2020)Jin, Covino, Liao, Li, D’Avanzo, Fan, \& Wei}]{jin2020kilonova}
Jin, Z.-P., Covino, S., Liao, N.-H., {et~al.} 2020, Nature Astronomy, 4, 77

\bibitem[{Kaplan {et~al.}(2020)Kaplan, McCandlish, Henighan, Brown, Chess, Child, Gray, Radford, Wu, \& Amodei}]{kaplan2020scaling}
Kaplan, J., McCandlish, S., Henighan, T., {et~al.} 2020, arXiv preprint arXiv:2001.08361

\bibitem[{Kluyver {et~al.}(2016)Kluyver, Ragan-Kelley, P{\'e}rez, Granger, Bussonnier, Frederic, Kelley, Hamrick, Grout, Corlay, {et~al.}}]{kluyver2016jupyter}
Kluyver, T., Ragan-Kelley, B., P{\'e}rez, F., {et~al.} 2016, in Positioning and power in academic publishing: Players, agents and agendas (IOS press), 87--90

\bibitem[{Liu {et~al.}(2021)Liu, Soria, Wu, Wu, \& Shang}]{liu2021sitian}
Liu, J., Soria, R., Wu, X.-F., Wu, H., \& Shang, Z. 2021, Anais da Academia Brasileira de Ci{\^e}ncias, 93, e20200628

\bibitem[{Loshchilov(2017)}]{loshchilov2017decoupled}
Loshchilov, I. 2017, arXiv preprint arXiv:1711.05101

\bibitem[{Loshchilov \& Hutter(2016)}]{loshchilov2016sgdr}
Loshchilov, I., \& Hutter, F. 2016, arXiv preprint arXiv:1608.03983

\bibitem[{Mart{\'\i}n {et~al.}(2015)Mart{\'\i}n, Ashish, Paul, Eugene, Zhifeng, Craig, Greg, Andy, Jeffrey, Matthieu, {et~al.}}]{martin2015tensorflow}
Mart{\'\i}n, A., Ashish, A., Paul, B., {et~al.} 2015, Software available from tensorflow. org, 7

\bibitem[{Mathur {et~al.}(2017)Mathur, Huber, Batalha, Ciardi, Bastien, Bieryla, Buchhave, Cochran, Endl, Esquerdo, {et~al.}}]{mathur2017revised}
Mathur, S., Huber, D., Batalha, N.~M., {et~al.} 2017, The Astrophysical Journal Supplement Series, 229, 30

\bibitem[{McInnes {et~al.}(2018)McInnes, Healy, \& Melville}]{mcinnes2018umap}
McInnes, L., Healy, J., \& Melville, J. 2018, arXiv preprint arXiv:1802.03426

\bibitem[{McKinney {et~al.}(2011)}]{mckinney2011pandas}
McKinney, W., {et~al.} 2011, Python for high performance and scientific computing, 14, 1

\bibitem[{Pan {et~al.}(2024{\natexlab{a}})Pan, Ting, \& Yu}]{pan2024astroconformer}
Pan, J.-S., Ting, Y.-S., \& Yu, J. 2024{\natexlab{a}}, Monthly Notices of the Royal Astronomical Society, 528, 5890

\bibitem[{Pan {et~al.}(2024{\natexlab{b}})Pan, Ting, Yu, Huang, \& Liu}]{pan2024scaling}
Pan, J.-S., Ting, Y.-S., Yu, J., Huang, Y., \& Liu, J.-F. 2024{\natexlab{b}}, in ICML 2024 AI for Science Workshop

\bibitem[{Paunzen {et~al.}(2024)Paunzen, Binder, Cyniburk, Duffek, Haberhauer, Heinreichsberger, Kohlhofer, Kue{\ss}, Maitzen, Saalmann, {et~al.}}]{paunzen2024apparent}
Paunzen, E., Binder, F., Cyniburk, A., {et~al.} 2024, arXiv preprint arXiv:2406.06174

\bibitem[{Pedregosa {et~al.}(2011)Pedregosa, Varoquaux, Gramfort, Michel, Thirion, Grisel, Blondel, Prettenhofer, Weiss, Dubourg, {et~al.}}]{pedregosa2011scikit}
Pedregosa, F., Varoquaux, G., Gramfort, A., {et~al.} 2011, the Journal of machine Learning research, 12, 2825

\bibitem[{Radford {et~al.}(2019)Radford, Wu, Child, Luan, Amodei, Sutskever, {et~al.}}]{radford2019language}
Radford, A., Wu, J., Child, R., {et~al.} 2019, OpenAI blog, 1, 9

\bibitem[{Rizhko \& Bloom(2024)}]{rizhko2024astrom}
Rizhko, M., \& Bloom, J.~S. 2024, arXiv preprint arXiv:2411.08842

\bibitem[{Roy {et~al.}(2024)Roy, Singh, Freitag, Schmude, Lal, Hegde, Ranjan, Lin, Gaur, Vos, {et~al.}}]{roy2024ai}
Roy, S., Singh, T., Freitag, M., {et~al.} 2024, arXiv preprint arXiv:2410.10841

\bibitem[{Sayeed {et~al.}(2021)Sayeed, Huber, Wheeler, \& Ness}]{sayeed2021swan}
Sayeed, M., Huber, D., Wheeler, A., \& Ness, M.~K. 2021, The Astronomical Journal, 161, 170

\bibitem[{Schafer(2011)}]{schafer2011savitzky}
Schafer, R.~W. 2011, IEEE Signal processing magazine, 28, 111

\bibitem[{Smith {et~al.}(2024)Smith, Roberts, Angeloudi, \& Huertas-Company}]{smith2024astropt}
Smith, M.~J., Roberts, R.~J., Angeloudi, E., \& Huertas-Company, M. 2024, arXiv preprint arXiv:2405.14930

\bibitem[{Tu {et~al.}(2022)Tu, Wu, Wang, Zhang, Liu, \& Wang}]{tu2022convolutional}
Tu, Z.-L., Wu, Q., Wang, W., {et~al.} 2022, The Astrophysical Journal, 935, 90

\bibitem[{Van~Cleve {et~al.}(2016)Van~Cleve, Howell, Smith, Clarke, Thompson, Bryson, Lund, Handberg, \& Chaplin}]{van2016s}
Van~Cleve, J.~E., Howell, S.~B., Smith, J.~C., {et~al.} 2016, Publications of the Astronomical Society of the Pacific, 128, 075002

\bibitem[{Wang {et~al.}(2024)Wang, Ge, Willis, Wang, Zhao, \& Hu}]{wang2024discovery}
Wang, K., Ge, J., Willis, K., {et~al.} 2024, Monthly Notices of the Royal Astronomical Society, 534, 1913

\bibitem[{Yang \& Liu(2019)}]{yang2019flare}
Yang, H., \& Liu, J. 2019, The Astrophysical Journal Supplement Series, 241, 29

\bibitem[{Yang {et~al.}(2017)Yang, Liu, Gao, Fang, Guo, Zhang, Hou, Wang, \& Cao}]{yang2017flaring}
Yang, H., Liu, J., Gao, Q., {et~al.} 2017, The Astrophysical Journal, 849, 36

\bibitem[{Yu {et~al.}(2020)Yu, Bedding, Stello, Huber, Compton, Gizon, \& Hekker}]{yu2020asteroseismology}
Yu, J., Bedding, T.~R., Stello, D., {et~al.} 2020, Monthly Notices of the Royal Astronomical Society, 493, 1388

\bibitem[{Yu {et~al.}(2018)Yu, Huber, Bedding, Stello, Hon, Murphy, \& Khanna}]{yu2018asteroseismology}
Yu, J., Huber, D., Bedding, T.~R., {et~al.} 2018, The Astrophysical Journal Supplement Series, 236, 42

\bibitem[{Yuan {et~al.}(2022)Yuan, Zhang, Chen, \& Ling}]{yuan2022einstein}
Yuan, W., Zhang, C., Chen, Y., \& Ling, Z. 2022, in Handbook of X-ray and Gamma-ray Astrophysics (Springer), 1--30

\bibitem[{Zhang {et~al.}(2024)Zhang, Helfer, Gagliano, Mishra-Sharma, \& Villar}]{zhang2024maven}
Zhang, G., Helfer, T., Gagliano, A.~T., Mishra-Sharma, S., \& Villar, V.~A. 2024, Machine Learning: Science and Technology

\bibitem[{Zuo {et~al.}(2024)Zuo, Tao, Liu, Xu, Zhang, Pan, Sun, Zhang, Cui, \& Yuan}]{zuo2024x}
Zuo, X., Tao, Y., Liu, Y., {et~al.} 2024, Research in Astronomy and Astrophysics, 24, 085016

\end{thebibliography}
\bibliographystyle{aasjournal}



\end{document}